\input phyzzx
\catcode`\@=11
\def\papers{\papersize\headline=\paperheadline\footline=\paperfootline}
\def\papersize{\hsize=6.5in \vsize=9in \hoffset=0in \voffset=0in
    \advance\hoffset by \HOFFSET \advance\voffset by\VOFFSET
    \pagebottomfiller=0pc
    \skip\footins=\bigskipamount \normalspace }
\catcode`\@=12
\papers
\overfullrule=0pt
\tolerance=5000
\normaldisplayskip = 20pt plus 5.0pt minus 10.0pt
\twelvepoint
\def\myspacing{\baselineskip=16pt}
\def\myabsspacing{\baselineskip=16pt}
\def\npb#1#2#3{{\it Nucl.\ Phys.} {\bf B#1} (19#2) #3}

\def\plb#1#2#3{{\it Phys.\ Lett.} {\bf B#1} (19#2) #3}
\def\prl#1#2#3{{\it Phys.\ Rev.\ Lett.} {\bf #1} (19#2) #3}

\def\prd#1#2#3{{\it Phys.\ Rev.} {\bf D#1} (19#2) #3}

\def\mpla#1#2#3{{\it Mod.\ Phys.\ Lett.} {\bf A#1} (19#2) #3}

\def\ijmp#1#2#3{{\it Int.\ J. Mod.\ Phys.} {\bf A#1} (19#2) #3}
\def\cmp#1#2#3{{\it Commun.\ Math.\ Phys.} {\bf #1} (19#2) #3}
\def\jgp#1#2#3{{\it J. Geom.\ Phys.} {\bf #1} (19#2) #3}

\def\endli{\hfill\break}
\def\frac#1#2{{#1 \over #2}}

\def\p{\partial}
\def\der{\,{\rm d}}

\def\e#1{\,{\rm e}^{#1}}

\def\ztwo{{\bf Z}_2}
\def\semi{\subset\kern-1em\times\;}

\def\su#1{{\rm SU}(#1)}
\def\u#1{{\rm U}(#1)}
\def\so#1{{\rm SO} (#1)}
\def\o#1{{\rm O} (#1)}
\def\sp#1{{\rm Sp} (#1)}

\def\ints{\int_{\Sigma}\der^2\sigma\;}

\def\rth{\sqrt{h}\;}
\def\rtg{\sqrt{g}\;}

         \def\Diff#1{{\rm Diff}(#1)}
      \def\Diffo#1{{\rm Diff}_0(#1)}
\def\christ#1#2#3{\Gamma^{#1}_{#2#3}\,}
              
\def\modspace#1{(M_G)^{#1}/S_{#1}}
                   \def\CC{{\cal C}}
\def\CD{{\cal D}}                   \def\CG{{\cal G}}
                   \def\CJ{{\cal J}}
\def\CL{{\cal L}}                   \def\CM{{\cal M}}
\def\CN{{\cal N}}                   \def\CO{{\cal O}}
\def\CR{{\cal R}}                   \def\CU{{\cal U}}
                   \def\CW{{\cal W}}
\def\CZ{{\cal Z}}
%
\pubnum{PUPT-1547}
\date{June 1995\cr hep-th/9507060\cr}
\titlepage
\title{Topological Rigid String Theory and Two Dimensional QCD}
\author{Petr Ho\v rava\foot{E-mail address: horava@puhep1.princeton.edu}}  
\medskip
\address{\centerline{Joseph Henry Laboratories}
\centerline{Jadwin Hall}
\centerline{Princeton University}
\centerline{Princeton, NJ 08544, USA}}
\bigskip
\abstract\myabsspacing\noindent
We present a string theory that reproduces the large-$N$ expansion of two 
dimensional Yang-Mills gauge theory on arbitrary surfaces.  First, a new 
class of topological sigma models is introduced, with path integrals 
localized to the moduli space of harmonic maps.  The Lagrangian of these 
harmonic topological sigma models is of fourth order in worldsheet 
derivatives.  Then we gauge worldsheet diffeomorphisms by introducing the 
induced worldsheet metric.  This leads to a topological string theory, whose 
Lagrangian coincides in the bose sector with the rigid string Lagrangian 
discussed some time ago by Polyakov and others as a candidate for QCD string 
theory.  The path integral of this topological rigid string theory is 
localized to the moduli spaces of minimal-area maps, and calculates their 
Euler numbers.  The dependence of the large-$N$ QCD partition functions on 
the target area emerges from measuring the volume of the moduli spaces, and 
can be reproduced by adding a Nambu-Goto term (improved by fermionic terms) 
to the Lagrangian of the topological rigid string.  
\endpage
\myspacing
\REF\dualbook{For a review, see e.g.: ``Dual Theory,'' ed.: M.~Jacob, 
{\it Phys.\ Rep.\ Reprint Series} {\bf 1} (North-Holland, 1974)\endli
P.H. Frampton, ``Dual Resonance Models and Superstrings'' (World Scientific, 
1986)}
\REF\nambu{Y. Nambu, ``Duality and Hadrodynamics,'' Chicago notes (1970), 
reprinted in:  ``Strings, Lattice Gauge Theory, High Energy Phenomenology,'' 
eds: V.~Singh and S.R.~Wadia (World Scientific, 1987)}
\REF\wilsonstr{K.G. Wilson, ``Confinement  of Quarks,'' \prd{10}{74}{2445}}
\REF\thooft{G. 't Hooft, ``A Planar Diagram Theory for Strong Interactions,'' 
\npb{72}{74}{461}}
\REF\qcdrev{J. Polchinski, ``Strings and QCD?,'' Austin preprint (June 1992)
\endli
D.J. Gross, ``Some New/Old Approaches to QCD,'' in: ``Integrable Quantum 
Field Theories,'' eds: L.~Bonora et al. (Plenum Press, 1993)}
\REF\seibw{N. Seiberg and E. Witten, ``Electric-Magnetic Duality, Monopole 
Condensation, and Confinement in $N=2$ Supersymmetric Yang-Mills Theory,'' 
\npb{426}{94}{19} (Erratum: \npb{430}{94}{485}); ``Monopoles, Duality and 
Chiral Symmetry Breaking in $N=2$ Supersymmetric QCD,'' \npb{431}{94}{484}
\endli
N. Seiberg, ``Electric-Magnetic Duality in Supersymmetric Non-Abelian Gauge 
Theories,'' hep-th/9411149}
\REF\thmodel{G. 't Hooft, ``A Two-Dimensional Model for Mesons,'' 
\npb{75}{74}{461}\endli
C.G. Callan, N. Coote and D.J. Gross, ``Two-Dimensional Yang-Mills Theory: 
A Model for Quark Confinement,'' \prd{13}{76}{1649}\endli
M.B. Einhorn, ``Confinement, Form Factors, and Deep-Elastic Scattering in 
Two-Dimensional Quantum Chromodynamics,'' \prd{14}{76}{3451}}
\REF\migdal{A.A. Migdal, ``Recursion Equations in Gauge Field Theories,'' 
{\it Zh.\ Eksp.\ Theor.\ Fiz.} {\bf 69} (1975) 810 [{\it Sov.\ Phys.\ JETP}
{\bf 42} (1975) 413]}
\REF\wittentwo{E. Witten, ``On Quantum Gauge Theories in Two Dimensions,'' 
\cmp{141}{91}{153}\endli
M. Blau and G. Thompson, ``Quantum Yang-Mills Theory on Arbitrary Surfaces,'' 
\ijmp{7}{92}{3781}}
\REF\rusakov{B.Ye.\ Rusakov, ``Loop Averages and Partition Functions in 
$\u N$ Gauge Theory on Two-Dimensional Manifolds,'' \mpla{5}{90}{693}}
\REF\wittenre{E. Witten,``Two Dimensional Yang-Mills Theory Revisited,'' 
\jgp{9}{92}{303}}
\REF\grfirst{D.J. Gross, ``Two Dimensional QCD as a String Theory,'' 
\npb{400}{93}{161}}
\REF\grta{D.J. Gross and W. Taylor, ``Two Dimensional QCD is a String 
Theory,'' \npb{400}{93}{181}; ``Twists and Wilson Loops in the String Theory 
of Two Dimensional QCD,'' \npb{403}{93}{395}}
\REF\kazkos{V.A. Kazakov and I.K. Kostov, ``Non-Linear Strings in Two 
Dimensional U($\infty$) Gauge Theory,'' \npb{176}{80}{199}\endli
V.A. Kazakov, ``Wilson Loop Average for an Arbitrary Contour in 
Two-Dimensional U($N$) Gauge Theory,'' \npb{179}{81}{283}}
\REF\trst{P. Ho\v rava, ``Topological Strings and QCD in Two Dimensions,'' 
hep-th/9311156, in: ``Quantum Field Theory and String Theory,'' eds: L. 
Baulieu et al., Proceedings of the Carg\`ese Workshop on ``New Developments 
in String Theory, Conformal Models, and Topological Field Theory,'' May 1993
(Plenum Press, 1995)}
\REF\future{P. Ho\v rava, in preparation}
\REF\polya{A.M. Polyakov, ``Fine Structure of Strings,'' \npb{268}{86}{406}; 
``Gauge Fields and Strings'' (Harwood Academic Publishers, 1987)}
\REF\kleinert{H. Kleinert, ``The Membrane Properties of Condensing Strings,'' 
\plb{174}{86}{335}}
\REF\otherrs{R.D. Pisarski, ``Perturbative Stability of Smooth Strings,'' 
\prl{58}{87}{1300}; ``Smooth Strings at Large Dimension,'' \prd{38}{88}{578}; 
``Heavy and Smooth Strings in QCD,'' in: ``String Theory -- Quantum Cosmology 
and Quantum Gravity; Integrable and Conformal Invariant Theories,'' eds: 
H.J.~de Vega and N.~S\'anchez (World Scientific, 1987)\endli
T.L. Curtright, G.I. Ghandour, C.B. Thorn and C.K. Zachos, ``Trajectories of 
Strings with Rigidity,'' \prl{57}{86}{799}\endli
T.L. Curtright, G.I. Ghandour and C.K. Zachos, ``Classical Dynamics of 
Strings with Rigidity,'' \prd{34}{86}{3811}\endli
E. Braaten and C.K. Zachos, ``Instability of the Static Solution to the 
Closed String with Rigidity,'' \prd{35}{87}{1512}\endli
E. Braaten, R.D. Pisarski and S.-M. Tse, ``Static Potential for Smooth 
Strings,'' \prl{58}{87}{93}\endli
P. Olesen and S.-K. Yang, ``Static Potential in a String Model with Extrinsic 
Curvatures,'' \npb{283}{87}{73}\endli
F. Alonso and D. Espriu, ``On the Fine Structure of Strings,'' 
\npb{283}{87}{393}}
\REF\bersh{M. Bershadsky, S. Cecotti, H. Ooguri and C. Vafa, ``Holomorphic 
Anomalies in Topological Field Theories,'' \npb{405}{93}{279}}
\REF\drudd{R.H.~Dijkgraaf and R.~Rudd, unpublished} 
\REF\cmr{S. Cordes, G. Moore and S. Ramgoolam, ``Large-$N$ 2D Yang-Mills 
Theory and Topological String Theory,'' Yale preprint (February 1994), 
hep-th/9402107}
\REF\tsm{E. Witten, ``Topological Sigma Models,'' \cmp{118}{88}{601}\endli
L. Baulieu and I.M. Singer, ``The Topological Sigma Model,'' 
\cmp{125}{89}{227}}
\REF\eellsrev{J. Eells and L. Lemaire, ``A Report on Harmonic Maps,'' {\it 
Bull.\ London Math.\ Soc.} {\bf 10} (1978) 1; ``Another Report on Harmonic 
Maps,'' {\it Bull.\ London Math.\ Soc.} {\bf 20} (1988) 385}
\REF\eellspaper{J. Eells and L. Lemaire, ``On the Construction of Harmonic 
and Holomorphic Maps Between Surfaces,'' {\it Math.\ Ann.} {\bf 252} (1980) 
27}
\REF\twistors{F.E. Burstall and J.H. Rawnsley, ``Twistor Theory for 
Riemannian Symmetric Spaces with Applications to Harmonic Maps of Riemann 
Surfaces,'' {\it L. N. Math.} {\bf 1424} (Springer, 1990)}
\REF\doubletop{M. Blau and G. Thompson, ``$N=2$ Topological Gauge Theory, 
the Euler Characteristic of Moduli Spaces, and the Casson Invariant,'' 
\cmp{152}{93}{41}}
\REF\vafaw{C. Vafa and E. Witten, ``A Strong Coupling Test of $S$-Duality,'' 
\npb{431}{94}{3}}
\REF\topomech{E. Witten, ``Supersymmetry and Morse Theory,'' {\it J. Diff.\ 
Geom.} {\bf 17} (1982) 661\endli
D. Friedan and P. Windey, ``Supersymmetric Derivation of the Atiyah-Singer 
Index and the Chiral Anomaly,'' \npb{235[FS11]}{84}{395}\endli
L. Alvarez-Gaum\'e, ``Supersymmetry and the Atiyah-Singer Index Theorem,'' 
\cmp{90}{83}{161}}
\REF\toptorus{P. Ho\v rava, ``Two Dimensional String Theory and the 
Topological Torus,'' \npb{386}{92}{383}; ``Spacetime Diffeomorphisms and 
Topological $w_\infty$ Symmetry in Two Dimensional Topological String 
Theory,'' \npb{414}{94}{485}}
\REF\etsm{P. Ho\v rava, ``Equivariant Topological Sigma Models,'' 
\npb{418}{94}{571}}
\REF\wittencs{E. Witten, ``Chern-Simons Gauge Theory as a String Theory,'' 
IAS preprint}
\REF\itoikub{C. Itoi and H. Kubota, ``Gauge Invariance Based on the Extrinsic 
Geometry in the Rigid String,'' {\it Z. Phys.} {\bf C44} (1989) 337; ``BRST 
Quantization of the String Model with Extrinsic Curvature,'' 
\plb{202}{88}{381}}
\REF\mathaiq{V. Mathai and D. Quillen, ``Superconnections, Thom Classes, and 
Equivariant Differential Forms,'' {\it Topology} {\bf 25} (1986) 85}
\REF\atiyahj{M.F. Atiyah and L. Jeffrey, ``Topological Lagrangians and 
Cohomology,''\jgp{7}{90}{119}}
\REF\ezra{N. Berline, E. Getzler and M. Vergne, ``Heat Kernels and Dirac 
Operators'' (Springer, 1992)}
\REF\blau{M. Blau, ``The Mathai-Quillen Formalism and Topological Field 
Theory,'' \jgp{11}{93}{95}}
\REF\fomenko{T.T. Dao and A.T. Fomenko, ``Minimal Surfaces, Stratified 
Multivarifolds, and the Plateau Problem,'' AMS Translations Vol.\ {\bf 84} 
(Providence, 1991)}
\REF\thurston{W. Thurston, ``The Geometry and Topology of Three-Manifolds ,'' 
Ch.\ 13, Princeton notes, 1977 (unpublished)}
\REF\wittenmir{E. Witten, ``Mirror Manifolds and Topological Field Theory,''  
in: ``Essays on Mirror Symmetry,'' ed.: S.-T. Yau (International Press, 1992)}
\REF\genqcd{O. Ganor, J. Sonnenschein and S. Yankielowicz, ``The String 
Theory Approach to Generalized 2D Yang-Mills Theory,'' Tel~Aviv preprint 
TAUP-2182-94 (July 1994)}
\REF\topograv{E. Witten, ``On the Structure of the Topological Phase of 
Two-Dimensional Gravity,'' \npb{340}{90}{281}; ``Two Dimensional Gravity and 
Intersection Theory on Moduli Space,'' {\it Surv.\ Diff.\ Geom.} {\bf 1} 
(1991) 243} 
\REF\othergrs{S.C. Naculich, H.A. Riggs and H.J. Schnitzer, ``Two-Dimensional 
Yang-Mills Theories are String Theories,'' \mpla{8}{93}{2223}\endli
S. Ramgoolam, ``Comment on Two Dimensional $\o N$ and $\sp N$ Yang-Mills 
Theories as String Theories,'' Yale preprint (1993)}
\REF\dualqcd{M. Douglas, ``Conformal Field Theory Techniques for Large $N$ 
Group Theory,'' hep-th/9303159\endli
R. Rudd, The String Partition Function for QCD on the Torus,'' hep-th/9407176}
\REF\rankdual{S.C. Naculich, H.A. Riggs and H.J. Schnitzer, ``Group-Level 
Duality in WZW Models and Chern-Simons Theory,'' \plb{246}{90}{417}}
\REF\stringdual{C.N. Hull and P. Townsend, ``Unity of String Dualities,'' QMW 
preprint (October 1994), hep-th/9410167\endli
E. Witten, ``String Theory Dynamics in Various Dimensions,'' IAS preprint 
IASSNS-HEP-95-18 (March 1995), hep-th/9503124}
\REF\barton{B. Zwiebach, ``Closed String Field Theory: Quantum Action and 
the B-V Master Equation, \npb{390}{93}{33}}
\REF\finiten{J. Baez and W. Taylor, ``Strings and Two-Dimensional QCD for 
Finite $N$,'' MIT preprint (January 1994), hep-th/9401041}
\REF\polchgr{M.B. Green and J. Polchinski, ``Summing over Worldsheet 
Boundaries,'' Santa Barbara preprint (June 1994), hep-th/9406012\endli
J. Polchinski, ``Combinatorics of Boundaries in String Theory,'' 
Santa barbara preprint (July 1994), hep-th/9407031\endli
M.B. Green, ``A Gas of D-Instantons,'' CERN preprint CERN-TH/95-78 (March 
1995)}
\chapter{Introduction}

Quantum string theory emerged from dual models [\dualbook,\nambu] as an 
attempt to formulate a theory of strong interactions directly in terms of 
degrees of freedom relevant at low energies.  Today, as in its early days, 
the string picture of hadrons and their interactions is based on experimental 
data (especially, the successes of Regge phenomenology).  

Ever since the discovery of quantum chromodynamics, intense efforts have been 
made to understand the theory better in the infrared, in particular to relate 
the successes of dual models with the intuitive picture of the color flux 
tube as the confining stringy force between quarks.  In QCD, this string 
picture of quark confinement has been further reinforced by results of the 
strong coupling expansion [\wilsonstr] (especially on the lattice, leading to 
the area law for Wilson loops), and the large-$N$ expansion with $N$ the 
number of colors [\thooft] (with its suggestive classification of diagrams in 
terms of surface topologies).  For recent reviews, see [\qcdrev].  

Despite these facts, the hypothetical QCD string theory has turned out to 
be surprisingly elusive.  The search for the string theory of quark 
confinement has produced very attractive and successful spin-offs (such as 
the string theory of quantum gravity), but a theory that captures adequately 
the relevant degrees of freedom in the confining regime of gauge theories 
still remains to be identified.  

Short of a theory of confinement that would be equivalent to standard 
QCD with bosonic gluons and three colors, one might try to first understand 
the problem in a simplified setting.  Two simplifications come to mind:  
(i) spacetime supersymmetry, and (ii) the large-$N$ limit.  The first option 
has been studied extensively in the past year or so, with some spectacular 
results [\seibw].  In this paper we follow the latter option.  Since 
the results of this paper suggest the existence of an $\CN=2$ supersymmetry 
on the worldsheet of the QCD string (at least in two target dimensions), one 
can even speculate that (i) and (ii) might be related, the link between them 
being the presence of supersymmetry.  

In two spacetime dimensions, the large-$N$ theory can be analyzed in detail 
[\thmodel], and does indeed lead to (perturbative) color confinement, as well 
as an infinite number of resonances when quarks are present in the 
microscopic theory.  These facts, together with the exact solvability of the 
Yang-Mills theory on arbitrary two-dimensional surfaces [\migdal-\wittenre], 
make the two dimensional theory an excellent starting point for the study of 
QCD string theory.  Since quarks play a relatively minor role in the string 
formation, we can as well study the pure Yang-Mills theory on arbitrary 
surfaces and try to reformulate it as a theory of strings.  This is exactly 
the strategy initiated by Gross and Taylor in [\grfirst,\grta] (for an 
earlier work, see also [\kazkos]).  While the authors of [\grta] were able to 
interpret the large-$N$ expansion in terms of counting specific maps from 
auxiliary two-dimensional surfaces to the spacetime manifold, they do not 
present a string Lagrangian that would reproduce the same results by path 
integral over worldsheet geometries.  The main goal of this paper is to 
present such a path integral framework.  

One more remark is in order.  The idea behind QCD string theory is to offer 
an alternative description of strong interactions, dual to the standard 
description in terms of gluons and quarks.  Such a duality can be valid 
either effectively in the infrared, or exactly at all length scales.  The 
latter scenario is certainly more appealing, and if true, would be much 
harder to establish.  In particular, it would be extremely interesting to 
see whether string theory can deal with the regimes where the naive string 
picture apparently breaks down (non-zero thickness of the color flux tube, 
parton behavior of amplitudes in the UV).  

An ``exact'' duality between strings and Yang-Mills theory could also be 
interesting for purely mathematical reasons.  The results of this paper 
provide a very simple example of such a mathematical interest.  We present an 
exact relation between the large-$N$ Yang-Mills theory on arbitrary two 
dimensional surface $M_G$ of genus $G$, and topological rigid string theory 
on $M_G$.  Rephrased in a mathematical language, our results establish a 
relation between cohomological properties of moduli spaces of flat $\su N$ 
connections on $M_G$ as $N\rightarrow\infty$, and cohomological properties of 
moduli spaces of minimal-area maps from auxiliary surfaces $\Sigma_g$ of 
genus $g$ to $M_G$.  Similar relations could be expected in higher 
dimensions if an exact duality between strings and Yang-Mills theory holds.  

This paper is organized as follows.  In the rest of \S{1}, results of the 
large-$N$ expansion of $\su N$ Yang-Mills theory [\grta] are reviewed, with a 
particular emphasis on the stringent constraints they pose on the 2D QCD 
string theory.  We also recall some general aspects of the 2D Yang-Mills 
theory, to be able to draw some analogies with the worldsheet theory later.  
In \S{2} we present a new class of topological sigma models with path 
integrals localized to moduli spaces of harmonic maps.  While \S{2} is 
devoted to general aspects of these ``harmonic topological sigma models,'' 
\S{3} discusses their partition functions in the case of two-dimensional 
targets.  In \S{4} we introduce a topological string theory, with path 
integrals localized to moduli spaces of minimal area maps, and show that this 
theory is a topological version of the rigid string theory.  The partition 
functions of this topological string theory calculate Euler numbers of the 
moduli spaces.  In \S{5} the discussion is specialized to two target 
dimensions, and the Euler numbers of all moduli spaces are computed 
explicitly.  These results are then related in \S{6} to the large-$N$ 
expansion of 2D QCD in the zero-tension limit.  The area dependence of the 
large-$N$ QCD partition functions at  non-zero tension is shown to come from 
measuring the volume of the moduli spaces, and a class of deformations of the 
topological rigid string by cohomology classes of the moduli spaces is 
discussed which reproduces this area dependence.  Since the simplest 
deforming term is essentially the Nambu-Goto Lagrangian, our string theory 
can be viewed as an alternative quantization of the Nambu-Goto Lagrangian.  
We conclude with some remarks in \S{7}.  

Results presented here have been reported some time ago [\trst].  The purpose 
of this paper is to give their more detailed and systematic exposition, and 
provide a reference point for our further investigations.  Even though the 
natural extension of the model to higher target dimensions is no longer a 
topological string theory [\future], two dimensions represent the particular 
case in which topological methods are sufficient for the solution of the 
model.  For this reason, we limit ourselves in this paper to the topological 
aspects of the theory, and will discuss the extension to higher dimensions 
elsewhere [\future].   

\smallskip\noindent
\undertext{Comparison to Previous Results}
\par\nobreak\smallskip\nobreak

Before embarking on the detailed presentation, it seems worth while putting 
our results into historical perspective.  

The string theory presented in this paper is a topological version of the 
bosonic rigid string theory as proposed some ten years ago by Polyakov 
[\polya], Kleinert [\kleinert] and others [\otherrs] as a candidate for QCD 
string theory.  In this respect, the results of this paper can be considered 
a confirmation of this long-standing proposal in two dimensions, the novelty 
perhaps being that one needs a specific worldsheet supersymmetry (or 
topological BRST symmetry) to establish -- at least in two space-time 
dimensions -- the correspondence between rigid strings and QCD.  

Another point worth stressing is that starting with the papers by Gross and 
Taylor [\grfirst,\grta], the search for the Lagrangian of the two-dimensional 
QCD string has its own and already rich history.  Shortly after the seminal 
work of Gross and Taylor, various proposals have been made as to the 
structure of the worldsheet Lagrangian capable of reproducing the explicit 
results of [\grta].  One of the first ideas was to use holomorphic 
topological sigma models to reconstruct the QCD string partition functions, 
an approach that 
has been suggested and worked out in detail for the genus-one amplitude by 
Bershadsky, Cecotti, Ooguri and Vafa in [\bersh], and independently by 
Dijkgraaf and Rudd [\drudd].  However, it remains unclear in this approach 
how to reproduce all the details of [\grta], in particular the so-called 
$\Omega$-points that occur for higher genus target spaces.  

Another approach has been suggested in [\trst] -- the idea is to study 
topological string theory based on the moduli spaces of minimal-area maps.  
This topological string theory has been called ``topological rigid string'' 
in [\trst], as the bosonic part of its worldsheet Lagrangian reproduces the 
bosonic rigid string Lagrangian of Polyakov and others [\polya,\kleinert,
\otherrs].  It was noticed in [\trst] that the partition functions of the 
topological rigid string theory calculate the orbifold Euler character of the 
moduli spaces of the minimal-area maps, and a conjecture has been made that 
the partition functions of Gross and Taylor calculate such Euler characters.  

The crucial missing link has been provided in the important work of Cordes, 
Moore and Ramgoolam [\cmr].  Among other things, the authors of [\cmr] first 
constructed a novel topological string Lagrangian, whose partition functions 
calculate the orbifold Euler characters of certain moduli spaces -- the 
so-called ``Hurwitz spaces''.  Even more importantly, it was actually proven 
in [\cmr] that (the chiral sectors of) the two-dimensional QCD partition 
functions do indeed calculate the Euler characters of the corresponding 
moduli spaces.  In this respect, the reader will find our calculation of the 
Euler characters of the moduli spaces of minimal-area maps in \S{5.3} very 
similar to the original calculation as presented first in [\cmr] for the 
Hurwitz moduli spaces -- the present author does not claim any credit for 
originality there.  One important point should be stressed however:  the 
conceptual similarity of the actual calculation of the partition functions 
in the topological string theory of [\cmr] and the topological rigid string 
theory does not mean that the two theories are manifestly equivalent.  
Indeed, even though the Hurwitz moduli spaces and the minimal-area moduli 
spaces are so similar, their detailed structure and the way they emerge in 
the path integral are still at least superficially different.  Thus, for 
example, the orientation-reversing tubes that follow from the geometrical 
interpretation of the $Omega$-points [\grta] are conjecturally related to 
subtle contact interactions in [\cmr].  On the other hand, in the topological 
rigid string theory such connecting tubes emerge at an equal level with 
superficially simpler configurations such as branchpoints (see \S{5.2}), 
since they represent yet another allowed property of a map that solves the 
extremal-area condition $\Delta x^\mu=0$ that defines the moduli spaces of 
the topological rigid string theory.   The fact that these solutions are 
unstable explains some interesting minus signs in the QCD partition 
functions.  Additional substantial differences between the two approaches 
become apparent when one tries to include the area dependence in the 
partition functions; while it is nototiously difficult to reproduce the 
leading area-dependent exponential term $e^{(n+\tilde n)A}$ in the approach 
of [\cmr], we will see in \S{6.2} that in the topological rigid string this 
term emerges from measuring the area of the corresponding minimal-area map.  
More detailed analysis reveals more conceptual differences: For example, the 
requirement of vanishing ghost-number anomaly of the worldsheet theory is 
realized differently in the two approaches -- the authors of [\cmr] ensure 
this property by introducing what they call co-fields, while in the 
topological rigid string this property follows from a certain symmetry 
between its ghosts and antighosts.  This suggests that the theory of Cordes, 
Moore and Ramgoolam might be related to the topological rigid string theory 
when some fields are integrated out in the former.  (For some further 
indications in favor of this suggestion, see also \S{2.2}.) 

In general, however, the question whether the two topological string theories 
proposed as candidates for two-dimensional QCD string theory are actually 
equivalent is a difficult problem which has not been resolved as yet, and 
therefore represents an important challenge for further study.  It appears 
that a detailed comparison of the two theories will require some new 
techniques in topological string theory, and is unfortunately beyond the 
scope of the present paper.  

\section{The Large-$N$ Expansion in Two-Dimensional QCD}

The path integral of the Yang-Mills gauge theory on an arbitrary two 
dimensional compact surface $M$ with fixed metric $g_{\mu\nu}$, as defined by 
the Yang-Mills Lagrangian 
$$\CL_{\rm YM}=-\frac{1}{4e_0^2}\int_M\der^2x\;\rtg{\rm Tr}\left(F^{\mu\nu}
F_{\mu\nu}\right),\eqn\eeaa$$
can be solved exactly [\migdal-\wittenre], for any compact gauge group 
$\CG$.  The exact partition function is given by
$$\CZ(G,\CG,e_0^2A)=\sum_R\left({\rm dim}\;R\right)^{2-2G}\e{-e_0^2A\,C_2(R)},
\eqn\eeab$$
where the sum goes over all irreducible representations $R$ of the gauge 
group, $G$ is the genus of the spacetime manifold $M$, $e_0$ is the gauge 
coupling constant, $A$ denotes the total area of $M$ in the fixed Riemann 
metric $g_{\mu\nu}$, and $C_2(R)$ is the quadratic Casimir of $R$.  Several 
methods have been used to obtain this result, among them an exact lattice 
formulation [\migdal] and the Duistermaat-Heckman localization in an 
associated topological Yang-Mills theory [\wittenre].  

In [\grta], the exact formula was further analyzed for $\CG=\su N$; in 
particular, the partition functions were expanded in the powers of $1/N$, and 
the resulting numerical coefficients were interpreted in terms of rules for 
counting specific classes of maps from auxiliary two-dimensional surfaces 
(string worldsheets) to the spacetime manifold $M$.  Schematically, the 
results of [\grta] can be summarized in the following formula,%
\foot{Throughout this paper, we will limit our discussion of large-$N$ 
QCD (and subsequently the QCD string theory) to $G>0$, in order to avoid the 
complications that lead to the Douglas-Kazakov phase transition on the 
sphere.}
$$\CZ(G,N,\lambda A)=\sum_{g=1}^\infty\frac{1}{N^{2g-2}}
\sum_{n,\tilde n=0}^\infty\e{-(n+\tilde n)\lambda A/2}
\sum_{i=0}^{2g-2-(n+\tilde n)(2G-2)}\left(\frac{\lambda A}{2}\right)^{i}
\omega^{g,G}_{i,n,\tilde n}.\eqn\eead$$
$\lambda\equiv e_0^2N$ is the combination of $e_0$ and $N$ that is kept fixed 
in the large $N$ limit, and $\omega^{g,G}_{i,n,\tilde n}$ are $A$-independent 
numbers, whose explicit form can be extracted from the results of [\grta] 
(see also \S{6} below).  

In the ``string'' interpretation of \eead , $1/N$ is the string coupling 
constant, and $g$ is the genus of the (not necessarily connected) string 
worldsheet, which covers $M$ outside a finite number of points, where 
specific singularities (such as branchpoints and collapsed handles) can 
emerge.  In this interpretation, $n$ and $\tilde n$ correspond to the number 
of sheets that cover the target in the orientation preserving and reversing 
sectors, respectively, and $\omega^{g,G}_{i,n,\tilde n}$ are interpreted as 
specific symmetry factors that essentially count the number of covering maps 
with the allowed types of singularities.  In this section we will only need 
the following properties of $\omega^{g,G}_{i,n,\tilde n}$: (i) they vanish 
identically unless either $n$ or $\tilde n$ is positive, (ii) they are 
invariant under the reversal of worldsheet orientation, 
$n\leftrightarrow\tilde n$, and (iii) not all of them are positive.  

The results of [\grta] thus provide a wealth of data, and we can learn some 
useful lessons about the string theory that is supposed to reconstruct these 
data by a path integral over worldsheet geometries.  A brief look at \eead\ 
indicates that such a string theory should be highly unusual, especially in 
comparison with the string theories of quantum gravity.%
\foot{Lacking a better terminology, we will call the string theory of 
quantum gravity the ``fundamental string theory'' henceforth.}
Here is a list of several basic properties of the large-$N$ expansion as 
summarized in \eead , which will thus serve as constraints on the QCD string 
Lagrangian:  

\item{1.} Folds of the maps from the worldsheet to the target are dynamically 
suppressed; i.e.\ they are either gauge artifacts, or their contribution is 
identically zero.  

\item{2.} Contributions to \eead\ from trivial homotopy classes of maps are 
zero.  In particular, for targets of genus higher than zero, there is no 
contribution from worldsheets of spherical topology.  

\item{3.} Unlike fundamental string theory, the QCD string spectrum contains 
neither the tachyon (massless in two dimensions) nor the graviton-dilaton 
multiplet.  In fact, the large-$N$ expansion indicates [\grta] that the 
physical spectrum of the one-string states in 2D QCD string theory contains 
exactly one bosonic string state for each non-trivial element of $\pi_1(M)$.  

\item{4.} The string theory should be invariant under the $\ztwo$ parity 
transformation that reverses worldsheet orientation while leaving the target 
intact.  We call a theory that respects this $\ztwo$ symmetry 
``non-chiral.''  

\item{5.} The theory should generate negative weights in the path integral, 
which suggests the presence of worldsheet fermions. 

\item{6.} At general values of the gauge coupling $\lambda$, the theory 
only depends on the target metric through its total dimensionless area 
$\lambda A$ (and the individual areas $\lambda A_i$ in case of a subdivision 
of the spacetime by Wilson lines).  The theory is invariant under a 
$w_\infty$ symmetry of area-preserving target diffeomorphisms, inherited 
from the $w_\infty$ symmetry of \eeaa .  

\item{7.} The terms exponential in $\lambda A$ behave as $(n+\tilde n)
\lambda A$, where $n$ and $\tilde n$ is the number of sheets of the cover 
that preserve and reverse orientation, respectively.  The exponential terms 
thus measure the total induced area of the worldsheet (as opposed e.g.\ to 
the degree of the map, the latter being proportional to $n-\tilde n$), and 
$\lambda$ is the physical tension of the string.  The rest of the area 
dependence takes the form of finite polynomials in $A$ for a given set of 
data ($g,G,n,\tilde n,\ldots$), with the top power of $A$ equal to 
$2g-2-(n+\tilde n)(2G-2)$.  

\item{8.} The $\lambda=0$ limit (i.e.\ the limit of zero target area/zero 
gauge coupling constant/zero string tension) should be described by a 
topological string theory.  In this limit, the path integral should not 
depend on the spacetime metric, and all physical correlation functions 
(including the $\lambda\rightarrow 0$ limit of all the partition functions) 
should be topological invariants, only depending on such general data as the 
genus of the worldsheet and the target.  

In this paper we present a string theory that fulfills these requirements 
in a very simple manner.  It turns out, in fact, that once we construct a 
relatively simple theory that satisfies these criteria, direct calculations 
reproduce the results of the two-dimensional QCD on arbitrary compact 
surfaces (i.e., the qualitative structure of \eead\ {\it and} the numerical 
values of $\omega^{g,G}_{i,n,\tilde n}$).  Heuristically, this is so because 
the number of topological invariants that can be defined and subsequently 
calculated as partition functions of a two dimensional string theory is quite 
limited; in other words, not so many ``string universality classes'' exist 
that respect all the symmetries summarized above.  This also suggests that 
other worldsheet Lagrangians might exist that describe the same ``string 
universality class'' (i.e.\ give the same set of partition and correlation 
functions),%
\foot{Since our results were announced in [\trst], a very interesting 
alternative approach to the 2D QCD string theory has been suggested and 
discussed by Cordes, Moore and Ramgoolam [\cmr].  Although different from 
the topological rigid string theory discussed in [\trst] and the present 
paper, the theory proposed in [\cmr] can be probably considered a different 
realization of the same string universality class, at least in the regimes 
analyzed in [\cmr] (i.e., in the chiral sector and/or at $\lambda=0$.)}
and the question is which realization of the string universality 
class of large-$N$ 2D QCD will prove most efficient for extensions to higher 
dimensions.  

Several remarks are in order:  

\item{1.}  Extension of the results to higher dimensions is of course the 
central motivation for the study of two dimensional QCD strings.  This 
program can only be considered successful if it provides some hints about the 
string description of QCD in higher dimensions.  The Lagrangian presented 
here suggests a particularly natural and non-trivial extension to higher 
dimensions.  This extension goes well beyond topological theory, and will be 
discussed elsewhere [\future].  

\item{2.} Even though the two dimensional QCD string theory is topological 
at zero gauge coupling/zero string tension, its extension to non-zero string 
tension goes beyond topological theory.  In two dimensions, we can 
still add a tension term to the topological rigid string, and evaluate its 
contribution to the path integral perturbatively in $\lambda$ using 
topological methods.  A better substantiation of this procedure would 
however require an apparatus that goes beyond the scope of this paper.  

\item{3.} Our approach to the problem of QCD string theory is 
phenomenological, i.e.\ instead of deriving the worldsheet theory directly 
from first principles (such as the Yang-Mills path integral), we compare the 
two theories at the level of their physical correlation functions (i.e., we 
try to construct another representant of the same string universality 
class).  Any ``microscopic'' derivation of the string theory from the 
Yang-Mills theory would be extremely valuable.  

\section{Some Results in Two-Dimensional Yang-Mills Theory}

Before proceeding to the definition of our string theory, we first summarize 
some aspects of two-dimensional Yang-Mills gauge theory, on arbitrary compact 
surfaces.  This summary will prove useful later, since throughout this paper 
we will be encountering worldsheet phenomena quite reminiscent of the pattern 
already established in the spacetime Yang-Mills theory.  It is by no means a 
review, and should rather serve as a reminder of those selected features 
of 2D Yang-Mills theory that are directly relevant to the present paper; for 
more details, the reader is referred to the original sources [\wittenre].  

Start with a topological Yang-Mills theory, whose basic BRST multiplet is 
given by
$$[Q,A_\mu]=\psi_\mu,\qquad\{Q,\psi_\mu\}=D_\mu\phi,\qquad[Q,\phi]=0.
\eqn\eeacca$$
Here $A\equiv A_\mu\der x^\mu$ is the gauge field, $\psi\equiv\psi_\mu\der 
x^\mu$ the fermionic topological ghost, $D_\mu$ the covariant derivative 
defined by $A$, $\phi$ a scalar ghost-for-ghost; all fields are in the 
adjoint representation of the gauge group $\CG$.  We will also write 
$F\equiv F_{\mu\nu}\der x^\mu\wedge\der x^\nu$ for the field strength of 
$A_\mu$.  

The standard construction of the theory uses the flatness condition $F=0$ as 
the gauge fixing condition.  In order to write a Lagrangian for this theory, 
we must introduce antighost multiplets (see [\wittenre] for details).  The 
Lagrangian is then written as a BRST commutator, 
$$\CL=\int_M\der^2x\;\rtg\left(F^2+{\rm ghost\ terms}\right),\eqn\eeaccb$$
and its path integral is localized to the moduli spaces of flat connections 
on $M$.  

Without spoiling the topological symmetry, one can deform the Lagrangian by 
terms that are formally BRST commutators.  One such term $\CL'$ of ghost 
number minus two has been found in [\wittenre]; its addition to the original 
Lagrangian,
$$\CL\rightarrow\CL+t_0\CL',\eqn\eeaccz$$
leads to the following dramatic consequences:

\item{1.} Even though the new term in the Lagrangian is a BRST commutator, 
it does change the partition functions of the theory, since it brings in some 
new components of the moduli spaces from the infinity in the space of all 
gauge connections.  

\item{2.} The antighost multiplets can be integrated out, and the whole 
theory can be written exclusively in terms of the fields contained in the 
basic BRST multiplet \eeacca .  After the integration over the antighost 
multiplets, the Lagrangian is roughly given by 
$$\CL=\frac{1}{t_0}\int_M\der^2x\;\rtg\left((D_\mu F^{\mu\nu})^2+
{\rm ghost\ terms}\right).\eqn\eeaccc$$

\item{3.} The deformed theory is now localized to the moduli spaces of all 
solutions to the Yang-Mills equations $D_\mu F^{\mu\nu}=0$, rather than the 
moduli spaces of flat connections.  The path integral gets contributions not 
only from the absolute minima of the Yang-Mills action \eeaa , but also from 
unstable solutions of \eeaa .  

The partition function of the physical Yang-Mills theory can be evaluated as 
a correlation function of a specific physical observable in the associated 
deformed topological Yang-Mills theory.  Local physical observables of the 
topological theory are given by BRST cohomology classes (equivariant with 
respect to the Yang-Mills gauge symmetry), and can be constructed from 
invariant polynomials on the Lie algebra of the gauge group, by evaluating 
the polynomials at the Lie-algebra valued ghost-for-ghost field $\phi$ of 
\eeacca .  These point-like observables $\CO^{(0)}$ can be used to construct 
non-local observables, which are BRST invariant only when integrated over a 
cycle on the spacetime manifold.  The non-local observables are related to 
the point-like ones by the BRST descent hierarchy, 
$$\der\CO^{(0)}=\{Q,\CO^{(1)}\},\qquad\der\CO^{(1)}=\{Q,\CO^{(2)}\}
\qquad\der\CO^{(2)}=0.\eqn\eeaccd$$
The simplest such observables are given by 
$$\CO^{(0)}_0={\rm Tr}\,(\phi^2),\qquad\CO^{(1)}_0=2\,{\rm Tr}\,(\phi\psi),
\qquad\CO^{(2)}_0=2\,{\rm Tr}\,(\phi F-\psi\wedge\psi).\eqn\eeacce$$
The statement that relates the physical Yang-Mills partition function to the 
correlation function of a physical observable in the topological theory 
can then be summarized in the following formula: 
$$\left\langle 1\right\rangle_{\rm physical\ YM}=\left\langle\exp\left\{
-\int_M\left(\phi F-\psi\wedge\psi\right)-e_0^2\int_M\rtg\phi^2\right\}
\right\rangle_{\rm topo.\ YM}\eqn\eeac$$
The two observables on the right hand side are indeed the $\CO^{(0)}_0$ and 
$\CO^{(2)}_0$ of \eeacce .  The $\psi\wedge\psi$ term in the exponential 
plays a relatively minor role, while the remaining part of the exponential 
term in \eeac\ is exactly the Yang-Mills Lagrangian, rewritten in a 
first-order form with the bosonic ghost-for-ghost $\phi$ playing the role of 
an auxiliary field.  The path integral on the right hand side of \eeac\ is 
the one of the deformed topological theory, as defined by \eeaccc .  For 
more details see [\wittenre].  

The striking analogies between the spacetime theory (i.e.\ the Yang-Mills 
path integral) and the worldsheet theory (the path integral of the 2D QCD 
string) that we will encounter later on provide yet another example of 
the well-documented ``as above, so below'' phenomenon of string theory, in 
which the existence of certain structures in spacetime (such as gravity, 
gauge invariance, supersymmetry, orbifolds, duality, etc.) is tied to the 
existence of analogous structures on the string worldsheet.  This heuristic 
principle of string theory works remarkably well, for reasons that still 
remain mostly mysterious.  

\chapter{Harmonic Topological Sigma Models}

In \S{1.1} we argued that in the zero tension limit, the string theory of 2D 
QCD should be topological.  Since apparently the only field that we can use 
to construct a worldsheet Lagrangian is the map $\Phi$ from worldsheet 
$\Sigma$ to the target manifold $M$, the first crude expectation is that the 
theory may be a certain form of a topological sigma model.  The theory must 
be parity invariant on the worldsheet, though, which invalidates the standard 
topological sigma model.  In that theory, the path integral is dominated by 
holomorphic maps [\tsm], which of course makes the theory chiral.  Moreover, 
for generic fixed worldsheet and target metrics on two-dimensional manifolds 
of higher genera, no holomorphic maps exist [\eellsrev], and the path 
integral will be either identically zero or may even have problems with 
topological invariance.  One could start with a chiral theory, and worry 
about the full non-chiral theory later (with such options as a coupling of 
two sectors with opposite chirality either directly or via an anomaly).  It 
seems much more natural, however, to start with a non-chiral theory from the 
beginning, which is the approach that we follow in this paper.  

Instead of starting with holomorphic maps, we therefore choose a different 
gauge-fixing codition for the topological sigma model, namely harmonicity of 
$\Phi$ with respect to fixed metrics on $\Sigma$ and $M$.  This condition has 
several important properties:

\item{1.}  It is manifestly non-chiral.

\item{2.}  The moduli space $\CM$ of harmonic maps contains as a subspace the 
moduli space of holomorphic maps; due to the $\ztwo$ chiral symmetry, $\CM$ 
contains all anti-holomorphic maps as well.

\item{3.}  In the case of two-dimensional targets, a deep mathematical theory 
exists [\eellsrev-\twistors] that guarantees the existence of at least one 
harmonic map for a generic choice of the worldsheet and target metric, the 
worldsheet and target genera, and the homotopy class of $\Phi$.  (For more 
details, see \S{3}.)

Although the ultimate interest of this paper is in the topological rigid 
string theory (as the 2D QCD string theory), we discuss the harmonic 
topological sigma models in some detail first, for the following two 
reasons:  

\item{1.}  The topological rigid string is directly related to the harmonic 
topological sigma model, by gauging worldsheet diffeomorphism symmetry.  
In this section, we will be able to explain and understand some features 
shared by the topological rigid string theory in the much simpler setting of 
topological sigma models, i.e.\ before the formulas become complicated by 
gauged worldsheet diffeomorphisms and dynamical worldsheet gravity.  

\item{2.}  Harmonic topological sigma models are interesting in their own 
right.  We will see below that they are related in several distinct ways to 
their holomorphic counterparts.  Furthermore, the theory of harmonic maps 
between manifolds is itself an intensely studied part of mathematics, with 
many important results and unexpected ramifications (see e.g.\ [\eellsrev] 
and references therein).  Hence, harmonic topological sigma models should be 
relevant to algebraic geometry, mirror symmetry and fundamental string 
theory.  

The topological sigma model [\tsm] is a theory of maps 
$\Phi:\Sigma\rightarrow M$ from a worldsheet $\Sigma$ with coordinates 
$\sigma^a$, $a=1,2$,  
\foot{Here we assume that $\Sigma$ is oriented and without a boundary; 
an extension to worldsheets with boudaries and crosscaps is discussed in 
\S{2.5}.}
to a target manifold $M$ with coordinates $x^\mu$, $\mu=1,\ldots, D$.  We 
assume that $M$ carries a fixed Riemann structure, defined by a metric with 
components $g_{\mu\nu}$.  For the purposes of the gauge fixing, we choose 
a fixed auxiliary metric $h_{ab}$ on $\Sigma$.  Notice that unlike in 
standard topological sigma models, we do not pick a complex structure in the 
target, nor do we use the worldsheet complex structure explicitly.  

The Riemann structures on $\Sigma$ and $M$ allow us to define the Laplacian 
on the maps $\Phi$ from $\Sigma$ to $M$.  In coordinates, we have
$$\Delta x^\mu\equiv h^{ab}\nabla_a\p_bx^\mu=h^{ab}\left(\p_a\p_bx^\mu
+\christ\mu\sigma\rho\p_ax^\sigma\p_bx^\rho-\christ cab\p_cx^\mu\right).
\eqn\eeae$$
Here $\christ cab$ and $\christ\mu\sigma\rho$ are the Christoffel symbols of 
the metric connection associated with $h_{ab}$ and $g_{\mu\nu}$ 
respectively, and $\nabla_a$ denotes the covariant derivative on 
$T^\ast\Sigma\otimes\Phi^{-1}(TM)$.  

To use the harmonicity condition
$$\Delta x^\mu=0\eqn\eeaf$$ 
as the gauge fixing condition in the topological sigma model, we first 
introduce the BRST multiplet that maps $x^\mu$ to their ghosts, 
$$[Q,x^\mu]=\psi^\mu,\qquad\{Q,\psi^\mu\}=0.\eqn\eeag$$
While the ghost fields are uniquely determined by the original fields of the 
theory, in our case $x^\mu$, the antighosts and their auxiliary fields 
are determined by the gauge fixing condition.  In the case of \eeaf , the 
antighosts $\chi^\mu$ and auxiliaries $B^\mu$ are sections of $\Phi^{-1}(TM)$:
$$\{Q,\chi^\mu\}=B^\mu,\qquad[Q,B^\mu]=0.\eqn\eeah$$
The gauge fixing condition is then implemented by the following choice 
of the Lagrangian: 
$$\CL=\{ Q,\frac{1}{r_0}\ints\rth\chi^\mu g_{\mu\nu}(\Delta x^\nu+\frac{1}{2}
\christ\nu\sigma\rho\chi^\sigma\psi^\rho-\frac{1}{2}B^\nu)\}.\eqn\eeai$$
The apparently non-covariant term with the explicit dependence on $\christ\mu
\sigma\rho$ is needed for spacetime diffeomorphism invariance.  While this 
term is standard in topological sigma models, we will see a natural 
explanation of its existence in this particular model later.  

Upon performing the BRST commutator in \eeai , the Lagrangian becomes 
$$\eqalign{&\quad\CL=\frac{1}{r_0}\ints\rth\left\{-\frac{1}{2}B^\mu 
g_{\mu\nu}B^\nu+B^\mu g_{\mu\nu}(\Delta x^\nu+\christ\nu\sigma\rho\chi^\sigma
\psi^\rho)-\Delta x^\mu g_{\mu\nu}\,\christ\nu\sigma\rho\chi^\sigma\psi^\rho
\right.\cr
&\left.{}-\chi^\mu g_{\mu\nu}\,\Delta\psi^\nu-R_{\mu\nu\sigma\rho}\,h^{ab}
\p_ax^\nu\p_bx^\sigma\,\chi^\mu\psi^\rho-\left(\frac{1}{4}R_{\mu\nu\sigma\rho}
+\frac{1}{2}g_{\lambda\kappa}\christ\lambda\mu\sigma\christ\kappa\nu\rho
\right)\chi^\mu\psi^\sigma\chi^\nu\psi^\rho\right\}.\cr}\eqn\eeaj$$
In this formula, the Laplacian $\Delta$ acting on $\psi^\mu$ is the covariant 
Laplacian on $\Phi^{-1}(TM)$, i.e.\ 
$$\eqalign{\Delta\psi^\mu=h^{ab}\nabla_a\nabla_b\psi^\mu&\equiv h^{ab}\left(
\frac{}{}\p_a\p_b\psi^\mu+2\christ\mu\sigma\rho\p_ax^\sigma\p_b\psi^\rho-
\christ cab\p_c\psi^\mu\right)\cr
&\qquad\qquad{}+\christ\mu\sigma\rho\Delta x^\sigma\psi^\rho+h^{ab}\p_a
x^\sigma\p_bx^\rho\,\psi^\nu\left(\frac{}{}R^{\mu}{}_{\rho\sigma\nu}+\p_\nu
\christ\mu\sigma\rho\right).\cr}\eqn\eeak$$
While $x^\mu$ and $B^\mu$ carry ghost number zero, $\psi^\mu$ and $\chi^\mu$ 
are of ghost number $+1$ and $-1$, respectively.  The ghost number generates 
a $\u 1$ symmetry of the theory; as we will se below, this symmetry is 
even preserved quantum mechanically, unlike in standard topological sigma 
models.  

$B^\mu$ is an auxiliary field and can be eliminated from the Lagrangian by 
using its equation of motion, 
$$B^\mu=\Delta x^\mu+\christ\mu\sigma\rho\chi^\sigma\psi^\rho,\eqn\eeal$$
which reduces the Lagrangian to 
$$\eqalign{\CL=&\frac{1}{r_0}\ints\rth\left\{\frac{1}{2}\Delta x^\mu\,
g_{\mu\nu}\,\Delta x^\nu-\chi^\mu\,g_{\mu\nu}\,\Delta\psi^\nu\right.\cr
&\qquad\qquad\qquad\left.{}-R_{\mu\sigma\rho\nu}\,h^{ab}\p_ax^\sigma\p_b
x^\rho\,\chi^\mu\psi^\nu-\frac{1}{4}R_{\mu\nu\sigma\rho}\chi^\mu\psi^\sigma
\chi^\nu\psi^\rho\right\}.\cr}\eqn\eeam$$
The simplicity of \eeam\ makes it one of the most useful expressions for the 
Lagrangian of the harmonic topological sigma model.  \eeam\ is BRST invariant 
under 
$$[Q,x^\mu]=\psi^\mu,\qquad\{Q,\psi^\mu\}=0\qquad\{Q,\chi^\mu\}=\Delta x^\mu
+\christ\mu\sigma\rho\chi^\sigma\psi^\rho.\eqn\eean$$
To demostrate the nilpotence of this BRST charge and the invariance of the 
Lagrangian, we have to use the equations of motion, so these properties are 
only valid on-shell.  

Sometimes it is convenient to keep the off-shell BRST symmetry by retaining 
the auxiliary fields $B^\mu$ explicitly, and rewrite the Lagrangian in a 
first order form.  Up to a total derivative, we indeed have%
\foot{From now on, we simplify notation by introducing ``$\cdot$'' to 
denote the scalar product in $TM$; thus, we write $v^\mu g_{\mu\nu}w^\nu
\equiv v\cdot w$ for any two vectors $v^\mu,w^\nu$ from $TM$.} 
$$\CL=\{Q,\frac{1}{r_0}\ints\rth\left[-h^{ab}\nabla_a\chi\cdot\p_bx+
\frac{1}{2}\chi^\mu g_{\mu\nu}(\christ\nu\sigma\rho\chi^\sigma\psi^\rho-
B^\nu)\right]\}.\eqn\eeao$$
The explicit evaluation of the BRST commutator then gives
$$\eqalign{\CL&=\frac{1}{r_0}\ints\left\{\vphantom{\frac{}{}}-h^{ab}\nabla_a
B\cdot\p_bx+h^{ab}\nabla_a\chi\cdot\nabla_b\psi-\frac{1}{2}B^2+R_{\mu\sigma
\rho\nu}h^{ab}\p_ax^\sigma\p_bx^\rho\chi^\mu\psi^\nu\right.\cr
&\qquad\vphantom{\int}\left.{}+\left(B^\mu-\Delta x^\mu\right)g_{\mu\nu}
\christ\nu\sigma\rho\chi^\sigma\psi^\rho-\left(\frac{1}{4}R_{\mu\sigma\nu\rho}
+\frac{1}{2}\christ\lambda\mu\nu g_{\lambda\kappa}\christ\kappa\sigma\rho
\right)\chi^\mu\psi^\nu\chi^\sigma\psi^\rho\right\}.\cr}\eqn\eeap$$
This first-order form is useful for the comparison with the topological rigid 
string theory, whose Lagrangian is most natural in a similar first-order 
form.  

\section{Symmetry Between Ghosts and Antighosts}

The harmonic topological sigma model enjoys a remarkable property that is by 
no means generic to all topological field theories:  The ghost and antighost 
fields $\psi^\mu$ and $\chi^\mu$ are both sections of the same bundle 
$\Phi^{-1}(TM)$ over the worldsheet.  This fact suggests the possibility of a 
hidden symmetry in the Lagrangian.  Indeed, it is easy to show that the 
Lagrangian is indeed symmetric under a bosonic $\u 1$ symmetry that mixes 
ghosts and antighosts, and is generated by
$$\eqalign{[J,\psi^\mu]&=\chi^\mu,\cr[J,x^\mu]&=0,\cr}\qquad
\eqalign{[J,\chi^\mu]&=-\psi^\mu,\cr[J,B^\mu]&=0.\cr}\eqn\eeaq$$
The interchange of ghosts and antighosts is the $\ztwo$ subgroup of this 
$\u 1$.  

The existence of this $\u 1$ symmetry allows us to define another BRST-like 
supersymmetry charge $\bar Q$, 
$$\eqalign{[\bar Q,x^\mu]&=\chi^\mu,\cr\{\bar Q,\psi^\mu\}&=-B^\mu,\cr}\qquad
\eqalign{\{\bar Q,\chi^\mu\}&=0,\cr[\bar Q,B^\mu]&=0.\cr}\eqn\eear$$
The two supercharges are of course related by $\bar Q=[J,Q]$.  They 
anticommute with each other and are both nilpotent,
$$\{Q,\bar Q\}=0,\qquad Q^2=0,\qquad\bar Q^2=0.\eqn\eeas$$
Notice that all the fields of the theory now fall into one irreducible 
multiplet of the extended supersymmetry algebra \eeas .  This is again 
reminiscent of the topological Yang-Mills theory, which can also be expressed 
(after a deformation that leads from the moduli spaces of flat connections to 
the moduli spaces of all solutions to Yang-Mills equations) in terms of the 
single BRST multiplet that only contains ghosts and no antighosts.  

Because of the symmetry between ghosts and antighosts, $Q$ should no longer 
play a preferred role in the Lagrangian.  We can indeed use the new 
supercharge $\bar Q$ to write the Lagrangian as a double commutator, 
$$\CL=\{Q,[\bar Q,F]\}.\eqn\eeat$$
Because $\bar Q$ carries ghost number minus one, $F$ is a bosonic function 
of ghost number zero.  It is easy to demonstrate that the following choice of 
$F$, 
$$F=-\frac{1}{2r_0}\ints\rth\left(h^{ab}\p_ax\cdot\p_bx-\chi\cdot\psi\right),
\eqn\eeau$$
when substituted in \eeat , reproduces the Lagrangian of the harmonic 
topological sigma model, \eeap .

The extremely simple form of $F$ has important consequences.  We will use 
\eeau\ elsewhere [\future] to draw some useful analogies between the harmonic 
topological sigma models and topological rigid strings on one hand, and some 
models studied in other areas of physics, such as the physics of polymers, 
disordered systems and stochastic quantization, on the other.  Already at 
this stage, however, we can use the double commutator expression for the 
Lagrangian to explain the existence of the peculiar $\christ\mu\sigma
\rho$-dependent term in the gauge fixing fermion of \eeai , which now comes 
from the manifestly covariant gauge fixing boson $F$ as a variation of 
$g_{\mu\nu}$ under $\bar Q$.  

\section{Relations to Holomorphic Topological Sigma Models}

Harmonic topological sigma models are related in several distinct ways to 
their holomorphic counterparts.  Since this line of thought is not directly 
related to the main aim of this paper, we will discuss this issue only 
briefly.  These relations are important because they place the theory of 
harmonic topological sigma models into a wider and much better understood 
context, and provide a different perspective of the models.  

The key to one such correspondence is the analogy between the formal 
structure of the harmonic topological sigma models and the properties of the 
topological Yang-Mills theory (as summarized in \S{1.2}).  In the previous 
subsection we have seen that the harmonic topological sigma model exhibits a 
symmetry between ghosts and antighosts, and all fields of the theory form an 
irreducible multiplet of an extended supersymmetry algebra.  In analogy with 
the Yang-Mills theory (cf.\ \S{1.2}), we can interpret this multiplet as a 
multiplet of fields and ghosts in another topological field theory (with a 
double-topological symmetry [\doubletop], cf.\ also [\vafaw]), and the 
Lagrangian of the harmonic topological sigma model as an effective Lagrangian 
in which all auxiliaries and antighosts of the double-topological field 
theory have already been integrated out.  The associated topological field 
theory is a double-topological holomorphic sigma model.  

The construction can be outlined as follows.  Consider $\CN=4$ supersymmetric 
sigma model on a hyper-K\"ahler manifold, for which we choose $T^\ast M$ 
where $M$ is complex.  (After a relation to the harmonic topological sigma 
model on $M$ is established, nothing depends on the auxiliary complex 
structure of $M$.)  A topological twist exists that turns two fermionic 
partners of the target coordinates into worldsheet scalars.  These two 
fermionic scalars play the role of topological ghosts of the 
double-topological BRST algebra, which contains two fermionic scalar 
nilpotent charges.  The two remaining fermions of the original $\CN=4$ 
supersymmetry now are one-forms on the worldsheet, and can be interpreted as 
anti-ghost fields of the double-topological symmetry.  We can eliminate them 
(toghether with their auxiliaries) from the Lagrangian by deforming the 
Lagrangian of the twisted $\CN=4$ supersymmetric sigma model $\CL$ by a term 
which violates the ghost number of the double-topological theory by two, 
$$\CL\rightarrow\CL+r_0\CL'.\eqn\eeaua$$
Here $\CL'$ is chosen such that the anti-ghosts and auxiliaries of the 
double-topological sigma model can be eliminated by their equations of 
motion, which set them equal to worldsheet derivatives of the remaining 
fields.  Schematically, if $\chi_z^I$ denotes one of the anti-ghost fields 
of the double-topological sigma model, its equation of motion will relate it 
to one of the ghost fields $\psi^I$:
$$\chi^I_z\propto\frac{1}{r_0}\p_z\psi^I.\eqn\eeaub$$
After the elimination of all auxiliaries and anti-ghosts (i.e.\ all fields 
that carry a non-trivial worldsheet tensor structure), we are left with an 
effective Lagrangian which only contains the target coordinates $x^\mu$, 
another bosonic field $B^\mu$, and the ghost fields $\psi^\mu,\chi^\mu$ of 
the double-topological symmetry, and turns out to be the Lagrangian of the 
harmonic topological sigma model.  

Notice several facts:  

\item{1.} This interpretation of $\CL$ as a deformed double-topological 
sigma model allows us to take the limit of $r_0\rightarrow 0$ in the 
harmonic topological sigma model, which is otherwise not well defined in 
\eeam .  

\item{2.}  Since the double-topological holomorphic sigma model is a twisted 
version of a $\CN=4$ supersymmetric sigma model, the construction outlined 
above indicates that in the theory of harmonic topological sigma models we 
are in fact dealing with a twisted $\CN=4$ theory in disguise (cf.\ 
[\vafaw]).  An analogous observation can be made in the case of the 
topological rigid string theory discussed in \S{4}.  

\item{3.}  There is a remarkable similarity between the double-topological 
sigma model and the topological theory used in [\cmr] for the construction of 
an alternative approach to 2D QCD string theory.  Hence, the relation between 
the double-holomorphic topological sigma models and harmonic topological 
sigma models might shed some light on the relation between the string 
theory of [\cmr] on one hand, and the topological rigid string theory on the 
other.  

Another relation between harmonic topological sigma models and their more 
thoroughly studied holomorphic counterparts arises as follows.  The theory 
of harmonic maps between manifolds is one of the simplest mathematical 
theories that allows for a twistor description [\twistors].  This twistor 
correspondence canonically identifies the moduli space of harmonic maps from 
$\Sigma$ to $M$, with a moduli space of holomorphic maps from $\Sigma$ to a 
different target $M'$, associated with $M$ by the twistor correspondence.  
(In many cases, the natural almost complex structure of $M'$ is not 
integrable, which is not an obstacle for the construction of holomorphic 
topological sigma models [\tsm].)  Typically, $M'$ is fibered over $M$, and 
holomorphic maps to $M'$ project to harmonic maps to $M$.  For example, a  
typical pair of manifolds related by this twistor correspondence is 
$$M=S^4,\qquad M'={\bf C}P^3.\eqn\eefefe$$
If extendable to the full topological sigma model, the twistor correspondence 
between $M$ and $M'$ would give a map between the harmonic topological sigma 
model on $M$ and its holomorphic counterpart with a different target, $M'$.  
This is somewhat reminiscent of mirror symmetry, which also relates two 
topological sigma models with different target manifolds.  

\section{Partition Functions}

By arguments standard in topological theories, the path integral of the 
harmonic topological sigma model is independent of the coupling constant 
$r_0$, and can be exactly calculated in the semiclassical approximation.  In 
the weak coupling limit of $r_0\ll 1$, the whole integral is localized to the 
locus of solutions to the gauge-fixing condition, i.e.\ to the moduli space 
$\CM$ of harmonic maps.  (Notice that $\CM$ is also the moduli space of all 
classical solutions to an associated bosonic problem, in this case the 
bosonic non-linear sigma model.  This is again quite analogous to a similar 
phenomenon in the Yang-Mills theory, cf.\ \S{1.2}.)  

Given a harmonic map $\Phi_0$, the next step is to evaluate the one-loop 
determinants of quantum fluctuations around $\Phi_0$.  In topological field 
theories, the bosonic part of this determinant cancels against its fermionic 
counterpart (at least up to a sign).  In our case, this cancellation is a 
consequence of the following BRST formula, 
$$[Q,\Delta x^\mu]=\Delta\psi^\mu-R^{\mu}{}_{\sigma\rho\nu}\,h^{ab}\p_a
x^\sigma\p_bx^\rho\psi^\nu-\christ\mu\sigma\rho\Delta x^\sigma\,\psi^\rho.
\eqn\eeav$$
When evaluated on a harmonic map, the last term on the right hand side 
vanishes, and we are left with the linearized equation of motion for the 
ghost.  The non-zero modes of the corresponding operators are thus related by 
the BRST transformation, and the fermionic one-loop determinant (almost) 
exactly cancels the bosonic determinant: 
$$\frac{\det'\left(\delta^\mu_\nu\Delta-R^\mu{}_{\sigma\rho\nu}\,h^{ab}\p_a
x^\sigma\p_bx^\rho\right)}{\det'^{1/2}\left\{\left(\delta^\mu_\nu\Delta-
R^\mu{}_{\sigma\rho\nu}\,h^{ab}\p_ax^\sigma\p_bx^\rho\right)^2\right\}}=
(-1)^\#.\eqn\eeaw$$
In this formula, $\Delta$ is the Laplacian on $\Phi^{-1}(TM)$ (since both 
$\psi^\mu$ and the quantum fluctuations $\delta x^\mu$ of $x^\mu$ are 
sections of $\Phi^{-1}(TM)$), the prime means that zero modes are omitted, 
and $\#$ denotes the number of negative eigenvalues of $\delta^\mu_\nu\Delta-
R^\mu{}_{\sigma\rho\nu}\,h^{ab}\p_ax^\sigma\p_bx^\rho$.  

The only remaining calculation is the integration over the zero modes, 
in particular the integral over the moduli spaces of harmonic maps.  The 
fermionic zero modes satisfy the linearized equation of motion in the 
harmonic background,
$$\Delta\psi_0^\mu-R^{\mu}{}_{\sigma\rho\nu}\,h^{ab}\p_ax^\sigma\p_bx^\rho
\,\psi_0^\nu=0.\eqn\eeax$$
In the theory of harmonic maps, this equation is known as the Jacobi 
equation.  The number of integrable solutions to this equation measures the 
dimension of the moduli space of harmonic maps, and the zero modes of 
$\psi^\mu$ form a basis of one-forms on the moduli spaces.  It is useful to 
decompose a given fermionic zero mode $\psi^\mu_0$ in a normalized basis 
$f^\mu_I(\sigma)$ (with $I=1,\ldots,{\rm dim}\,\CM$) of solutions to \eeax , 
as follows:
$$\psi^\mu_0=\sum_Ia^If^\mu_I(\sigma).\eqn\eeaxxa$$
The anticommuting coefficients $a^I$ can be identified with one-forms on 
the moduli space, $a^I\propto\der m^I$, with $m^I$ a coordinate system on 
$\CM$.  

Since the anti-ghost zero modes $\chi^\mu_0$ satisfy exactly the same 
equation \eeax\ as the ghost zero modes $\psi^\mu_0$, they can be similarly 
written as
$$\chi^\mu_0=\sum_Ib^If^\mu_I(\sigma).\eqn\eeaxxb$$
Thus, the ghost number is preserved quantum mechanically, without any 
anomaly, and the partition functions can be non-zero without any insertions 
of BRST invariant observables.  

Since the curvature-dependent two-fermi term in $\CL$ was actually a 
part of the kinetic term of the fermions, the only term left in the zero-mode 
integration is the curvature-dependent four-fermi term, whose form is 
identical to the analogous four-fermi term in topological mechanics 
[\topomech].  Hence, the whole path integral reduces to
$$\int_\CM(-1)^\#\int\prod\der a^I\der b^I\exp\left\{-\frac{1}{4}a^Ia^Jb^Kb^L 
R_{IJKL}(m)\right\},\eqn\eeaxxc$$
where $R_{IJKL}$ is a tensor on the moduli space, induced from the target 
curvature tensor by
$$R_{IJKL}\equiv\ints\rth R_{\mu\nu\sigma\rho}f_I^\mu f_J^\nu f_K^\sigma 
f_L^\rho.\eqn\eeaxxd$$
The integral over the anti-ghost zero modes $b^I$ in \eeaxxc\ gives the 
Euler density on the moduli space, with the zero modes of $\psi^\mu$ playing 
the role of one-forms on $\CM$, $a^I\propto\der m^I$.  The remaining integral 
over $\CM$ thus gives the Euler number $\chi(\CM)$.  

In the homotopically trivial sector, the calculation is identical to that of 
topological mechanics.  The moduli space is the target manifold itself, the 
fermionic zero-mode integral gives the Euler character density on the target, 
and the integral over the bosonic moduli is equal to the Euler number 
$\chi(M)$.  Path integrals in non-trivial homotopy classes yield interesting 
stringy corrections to $\chi(M)$; in the simplest cases, they count the 
number of harmonic maps in the given homotopy class (cf.\ \S{3} below).  

\section{Observables and Correlation Functions}

Observables in topological field theories are defined as cohomology classes 
of the BRST charge.  In topological sigma models, the simplest observables 
are point-like on the worldsheet, and are in one-to-one correspondence with 
the cohomology classes of the target manifold.  Given a differential $s$-form 
$A\equiv\sum A_{\mu_1\ldots\mu_s}\der x^{\mu_1}\wedge\ldots\wedge{\rm d} 
x^{\mu_s}$ on $M$, we define
$$\CO_A=\sum A_{\mu_1\ldots\mu_s}\psi^{\mu_1}\ldots\psi^{\mu_s}.\eqn\eeay$$
$\CO_A$ is of course BRST invariant for $A$ a closed form, and BRST exact 
for $A$ an exact form.  Since this argument depends neither on the choice 
of the gauge fixing condition nor on the antighost multiplet, these 
``homology observables'' exist in any topological sigma model, independently 
of the specific choice of the gauge fixing condition.  

When the fundamental group of the target is non-trivial, another class of 
observables can exist.  For every element $\gamma$ of the fundamental group 
$\pi_1(M)$, consider the vacuum state $\CO_\gamma$ of the string in the 
winding sector $\gamma$.  This state is a non-trivial BRST cohomology class 
of the theory, as can be seen from the following argument.  The on-shell 
BRST transformation rules 
$$[Q,x^\mu]=\psi^\mu,\qquad\{Q,\psi^\mu\}=0,\qquad\{Q,\chi^\mu\}=\Delta 
x^\mu+\christ\mu\sigma\rho\chi^\sigma\psi^\rho\eqn\eeayy$$
indicate that states with $\chi^\mu=\psi^\mu=0$ are BRST invariant if they 
are annihilated by $\Delta x^\mu$, which means that the linear part of 
$x^\mu$ is BRST closed but not exact, and $\CO_\gamma$ are physical.  Since 
these new observables are parametrized by the elements of the first homotopy 
group of the target, it is natural to call them ``homotopy observables.''  
(Similar observables have been discussed in [\toptorus].)  The full space of 
physical observables is roughly a tensor product of the homotopy and homology 
sectors.  One must be careful, however, since the products of homology and 
homotopy observables can in some cases be singular, and the space of physical 
states is then a subspace in the direct product.  

Although the structure of the point-like observables in the harmonic and 
holomorphic topological sigma models is very similar, their correlation 
functions differ dramatically.  The symmetry between ghosts and antighosts, 
not shared by holomorphic topological sigma models, leads to the quantum 
mechanical conservation of the ghost number.  This absence of ghost number 
anomaly shows up in correlation functions as a special selection rule.  
Indeed, the correlation functions are zero, unless the total ghost number of 
all observables under the correlator vanishes:  
$$\left\langle\CO_1\ldots\CO_s\right\rangle=0\qquad{\rm if}\qquad\sum_{i=1}^s
\,{\rm ghost\ }\#\;(\CO_i)\neq 0.\eqn\eeaya$$
Since all point-like cohomology observables \eeay\ with $s>0$ carry a 
positive ghost number, all their correlations vanish identically:
$$\left\langle\CO_{A_1}\ldots\CO_{A_s}\right\rangle=0\qquad{\rm if\ deg}\,
(A_i)\neq 0\ {\rm for\ any}\ A_i.\eqn\eeayb$$
Unless we compensate for the ghost number of the non-trivial cohomology 
classes by additional insertions of (non-local) observables with negative 
ghost numbers, the physical observables corresponding to the cohomology 
classes of non-zero degree effectively decouple from the correlation 
functions, and we are left with the bosonic homotopy observables parametrized 
by the elements of $\pi_1(M)$.  This is qualitatively in a very good 
agreement with the results of the large-$N$ expansion in 2D QCD, where we 
have one physical string state for each non-trivial element of $\pi_1(M)$.  
In the topological rigid string, the trivial homotopy class becomes 
unphysical by virtue of worldsheet diffeomorphism invariance.  

\section{Equivariant Harmonic Topological Sigma Models}  

So far we have assumed that the worldsheet manifold is oriented and without 
a boundary.  Topological sigma models can be naturally extended to 
worldsheets with boundaries and/or crosscaps [\etsm].  While this extension 
may be interesting for pure mathematical reasons (since it refines the 
topological invariants already calculated by the theory on oriented 
surfaces), our motivation here is different.  Indeed, we conjecture in \S{7} 
that the equivariant theory is related to the Yang-Mills theory with 
alternative gauge groups ($\so N,\sp N$) in the same way the original theory 
is related to Yang-Mills theory with gauge group $\su N$.    

The framework that allows us to construct this extension systematically is 
the theory of equivariant topological sigma models.  The central idea is to 
consider the $\ztwo$ symmetry $I_0$ that reverses worldsheet orientation, 
i.e.\ acts in a suitable coordinate system $(z,\bar z)$ on $\Sigma$ by 
$$I_0:\Sigma\rightarrow\Sigma,\qquad I_0(z,\bar z)=(\bar z,z),\eqn\eeoo$$
and treat $I_0$ as an orbifold symmetry (i.e., a discrete gauge symmetry) of 
the theory on closed oriented surfaces.  The resulting orbifold model is a 
topological sigma model on worldsheets with boundaries/crosscaps.  

In order to construct an orbifold theory wit $I_0$ as a generator of the 
orbifold group, we must extend $I_0$ to a $\ztwo$ symmetry of the full 
quantum theory.  Since all fields of a topological sigma model are tensors of 
a specific degree on the worldsheet, a canonical extension of $I_0$ to the 
fields exists, which leaves the target intact.  If this canonical extension 
of $I_0$ were a symmetry of the model, it could be used to define a canonical 
orbifold theory (essentially, with standard, Neumann boudary conditions on 
the open strings).  

In holomorphic topological sigma models, this canonical extension of $I_0$ 
fails to be a symmetry of the quantum theory, since it transforms holomorphic 
maps into anti-holomorphic ones.  Consequently, there is no canonical 
orbifold theory parity associated with a given holomorphic topological sigma 
model (unless there are no non-trivial instantons, a case which is not of 
great interest).  To promote $I_0$ to a $\ztwo$ symmetry, we must pick an 
anti-holomorphic involution of the target, 
$$I:M\rightarrow M\eqn\eeooa$$
and gauge the diagonal $\ztwo$, 
$$\ztwo=\left\{1,I_0I\right\}.\eqn\eeoob$$
Twisted states of such an orbifold theory are open strings with both ends 
fixed to the submanifold of fixed points of $I$.  This class of models has 
been discussed in detail in [\etsm] (see also [\wittencs]).  Physical 
correlation functions of this theory calculate an equivariant extension of 
the quantum cohomology algebra on $M$.  

Harmonic topological sigma models, on the other hand, are by construction 
parity invariant on the worldsheet, and the canonical extension of $I_0$ 
is a symmetry of the theory and can be gauged.  Hence, a canonical orbifold 
model exists, with standard Neumann boundary conditions at both ends of the 
open strings.  Since these boudary conditions do not spoil the conservation 
of the ghost number, all homology observables with non-zero degree decouple 
from the theory by the same argument as in the closed string model.  Notice 
that the open-string sector does not produce any non-trivial homotopy 
observables, since every open string is homotopically trivial.  

\chapter{Harmonic Topological Sigma Models in Two Dimensions}

Since the definition of harmonic topological sigma models has not required a 
choice of a complex structure on the target, this class of topological sigma 
models can be studied in arbitrary target dimensions.  With our primary 
motivation in mind, however, we will now restrict our attention to 
two-dimensional targets.  

\section{Targets of Genus $G>1$}

Consider the path integral of the harmonic topological sigma model in a 
fixed homotopy class $[\Sigma_g,M_G]$ of maps from a two-dimensional 
worldsheet $\Sigma_g$ of genus $g$ to a two-dimensional target $M_G$ of genus 
$G$ with $G,g$ greater than one, and assume that the degree of the map is 
non-zero.  In the set of all homotopy classes, such choice is generic.  Given 
a fixed Riemannian metrics $h_{ab}$ on the worldsheet $\Sigma_g$ and 
$g_{\mu\nu}$ on the target $M_G$, a deep mathematical theorem [\eellsrev] 
proves the existence of exactly one harmonic map in $[\Sigma_g,M_G]$; 
moreover, this harmonic map minimizes the action of the bosonic sigma model 
within $[\Sigma_g,M_G]$.  Consequently, the one-loop determinants \eeaw\ 
cancel exactly, and we obtain an extremely simple answer for the partition 
function of the harmonic topological sigma model in such a (generic) homotopy 
sector:
$$\CZ([\Sigma_g,M_G])=1.\eqn\eeazab$$
Since the moduli space consists of only one point, the path integral 
trivially confirms the expectation that $\CZ$ is equal to the Euler number 
of the moduli space.  

Even in the homotopy classes of degree zero, harmonic maps always exist, and 
the worldsheet is mapped by any harmonic map either to a simple geodesic 
curve in $M_G$ (in case the homotopy class is non-trivial), or to a point in 
$M_G$ (in the trivial homotopy class).  In the latter case, the moduli space 
coincides with the target itself, and the partition function equals the Euler 
number of $M_G$.  

\section{Targets of Genus $G\leq 1$}

Since for targets of genus greater than one we are always guaranteed the 
existence of a harmonic map in any homotopy class, the path integral is 
always well-defined and easy to calculate.  General theory of harmonic maps 
between surfaces shows, however, that the situation is much more complicated 
for targets of low genera.  

Targets of genus one (with a flat metric $g_{\mu\nu}$) are easy to deal with, 
since the existence theorem is still valid.  A given harmonic map is no 
longer unique in its homotopy class, however, since we can use the target 
isometry group to generate a two-parameter class of harmonic maps 
parametrized by the target itself.  The evaluation of the path integral is 
then proportional to the Euler number of the target, and the partition 
function is always zero.  

When the target is either the sphere or the projective plane, there are known 
homotopy classes of maps with no harmonic representative, which makes the 
formal semiclassical evaluation of the path integral in such homotopy classes 
ill-defined.  Thus, for example, there is no harmonic map of degree $\pm 1$ 
from the torus to the sphere, no matter what metrics we choose on $\Sigma_g$ 
and $M_G$.  This singular behavior on the sphere is reminiscent of some 
properties of the large-$N$ QCD on the sphere (such as the Douglas-Kazakov 
phase transition).  A closer examination of the harmonic topological sigma 
model in these singular cases should certainly be interesting.  

\section{From Sigma Models to String Theory}

Before we go on and construct a topological string theory using the harmonic 
topological sigma model as worldsheet matter, let us briefly consider our 
options.  

In the topological sigma models, the topology of $\Sigma$ and the homotopy 
class of $\Phi$ have been fixed.  In string theory, we must be able to sum 
over all worldsheet genera and all homotopy classes of maps.  While trying to 
define this sum, we must deal with worldsheet diffeomorphisms.  Even though 
the partition functions of the harmonic topological sigma model in a given 
homotopy class are $\Diffo{\Sigma_g}$ invariant, the naive sum of the 
partition functions over all worldsheet genera and homotopy classes of maps, 
$$\sum_g g_{\rm string}^{2g-2}\sum_{[\Sigma_g,M_G]}\e{-c\;{\rm deg}\,\Phi}
\left\langle 1\right\rangle_{[\Sigma_g,M_G]},\eqn\eeaza$$
is infinite.  (In \eeaza , $g_{\rm string}$ is the string coupling constant, 
$\langle 1\rangle_{[\Sigma_g,M_G]}$ is the partition function of the harmonic 
topological sigma model in homotopy class $[\Sigma_g,M_G]$, ${\rm deg}\,\Phi$ 
is the degree of a map $\Phi\in[\Sigma_g,M_G]$, and $c$ is a ``chemical 
potential'' introduced here to regularize the sum over all values of ${\rm 
deg}\,\Phi$.)  This infinity can be 
easily traced back to the symmetry under global worldsheet diffeomorphisms, 
given by the mapping class group $\Gamma_{\Sigma_g}\equiv\Diff{\Sigma_g}/
\Diffo{\Sigma_g}$.  

There are several possible remedies for this infinity.  The minimal way which 
makes the sum over all homotopy classes finite is to simply divide the 
infinite sum by the (infinite) volume of the mapping class group, and define 
the string partition function by
$$\CZ=\sum_g g_{\rm string}^{2g-2}\frac{1}{{\rm Vol}\,(\Gamma_{\Sigma_g})}
\sum_{[\Sigma_g,M_G]}\e{-c\;{\rm deg}\,\Phi}\left\langle 1\right\rangle_{[
\Sigma_g,M_G]},\eqn\eeaz$$
Since we have already calculated earlier in this section most of the 
ingredients of the right hand side of \eeaz\ (most of the contributions are 
actually equal to one), direct comparison with \eead\ shows that the final 
result for $\CZ$ is indeed very different from the results of the large-$N$ 
expansion in 2D QCD.  

The second option is a little more sophisticated, since it makes the theory 
diffeomorphism invariant by coupling the matter theory to topological 
gravity.  While such theory in general depends on precise details of this 
coupling, the most straightforward approach is to consider harmonic maps and 
allow the worldsheet metric to vary.  This ``minimal'' coupling leads to a 
theory localized to moduli spaces which are canonically fibered over the 
moduli spaces of Riemann surfaces.  The minimal coupling leads to a further 
ramification, depending on how we treat the symmetry between ghosts and 
antighosts found in the harmonic topological sigma model.  We can 
either treat this symmetry as accidental and couple the matter to usual 
topological gravity, or we can interpret the harmonic topological sigma model 
as a theory with double topological symmetry and couple it to double 
topological gravity (a theory which calculates the Euler numbers of the 
moduli spaces of Riemann surfaces).  Although both of these conservative 
approaches might be of some independent interest and apparently lead to 
self-consistent theories, in this paper we follow a different route, 
explained in the following section.  

\chapter{Topological Rigid String Theory}

In the standard setting, the Lagrangian of a topological (string) theory is 
constructed as an exact BRST commutator,
$$\CL=\{Q,\Psi\},\eqn\eeba$$
where $\Psi$ is a suitably chosen gauge-fixing fermion.  In our case, 
the only fields that describe the string are the coordinates $x^\mu$ 
of the map $\Phi$ from the worldsheet $\Sigma$ to the spacetime $M$.  
In particular, we do not introduce an independent worldsheet metric, and will 
use the induced one whenever a metric is needed.  The basic BRST multiplet 
then consists of $x^\mu$ and their ghost partners $\psi^\mu$,
$$[Q,x^\mu]=\psi^\mu,\qquad\{Q,\psi^\mu\}=0.\eqn\eebb$$
Of course, $\psi^\mu$ are components of a section of $\Phi^{-1}(TM)$.  

As always in topological field theory, symmetries are more important than 
the Lagrangian itself, and we will discuss them first.  In addition to the 
topological symmetry we consider worldsheet diffeomorphisms a gauge 
symmetry.  This additional symmetry distinguishes the model from a theory of 
topological worldsheet matter and makes it a string theory.  

The presence of diffeomorphism invariance as an additional gauge symmetry 
requires new ghosts in the BRST multiplet, which now becomes 
$$[Q,x^\mu]=\psi^\mu+c^a\p_ax^\mu,\qquad\{Q,\psi^\mu\}=c^a\p_a\psi^\mu,\qquad
\{Q,c^a\}=c^b\p_bc^a,\eqn\eebc$$
and causes a typical overcounting of gauge symmetries.  As a consequence of 
this overcounting, the theory will enjoy a new, fermionic gauge symmetry 
given by 
$$\delta_\epsilon x^\mu=0,\qquad\delta_\epsilon\psi^\mu=\epsilon^a\p_ax^\mu.
\eqn\eebd$$
The standard strategy for taking care of this ghostly symmetry is to 
introduce a ghost for ghost field $\phi^a$, and extend the BRST multiplet to
$$\eqalign{[Q,x^\mu]&=\psi^\mu+c^a\p_ax^\mu,\cr\{Q,c^a\}&=c^b\p_bc^a-\phi^a,
\cr}
\qquad\eqalign{\{Q,\psi^\mu\}&=c^a\p_a\psi^\mu+\phi^a\p_ax^\mu,\cr
[Q,\phi^a]&=c^b\p_b\phi^a+\phi^b\p_bc^a.\cr}\eqn\eebe$$
The BRST multiplet is already becoming complicated, and we can simplify 
things by agreeing to work directly with diffeomorphism invariant 
configurations only.  This restriction allows us to ignore the diffeomorphism 
ghosts $c^a$, and leads to the so-called equivariant BRST quantization.  In 
fact, this equivariant approach will turn out to be very effective for the 
comparison of the topological rigid string to the large-$N$ expansion of 
two-dimensional QCD, as the moduli spaces emerging in the latter are 
manifestly diffeomorphism invariant.  In the equivariant theory, the BRST 
multiplet is reduced to 
$$[Q,x^\mu]=\psi^\mu,\qquad\{Q,\psi^\mu\}=\phi^a\p_ax^\mu,\qquad[Q,\phi^a]=0,
\eqn\eebee$$
and the BRST charge is only nilpotent on diffeomorphism invariant 
configurations.  

We will actually go one step further, and throughout most of the paper will 
keep the fermionic gauge symmetry along with the ordinary diffeomorphism 
invariance as a manifest gauge symmetry of the theory, without explicitly 
fixing either of them.  One of the benefits of this strategy is the 
simplification of the subsequent formulas, which would otherwise contain many 
terms depending on $\phi^a$ and the gauge-fixing multiplets associated with 
it.  

To construct a Lagrangian, we need a gauge fixing condition; we choose the 
minimal-area condition, 
$$\Delta x^\mu=0.\eqn\eebf$$
The Laplacian in \eebf\ is now the covariant Laplacian on $x^\mu$, defined 
with respect to the induced metric on the worldsheet:%
\foot{Hoping not to create too much confusion, we keep the notation of 
the previous section.  In this section, the worldsheet metric $h_{ab}$ is 
always the induced metric, while in the previous section, it was always the 
fixed auxiliary metric.}
$$\Delta x^\mu\equiv h^{ab}\nabla_a\p_bx^\mu=h^{ab}\left(\delta^\mu_\nu-
\p_cx^\mu h^{cd}\p_dx^\lambda g_{\lambda\nu}\right)\left(\p_a\p_b
x^\nu+\christ\nu\sigma\rho\p_ax^\sigma\p_bx^\rho\right).\eqn\eebg$$
This of course means that the maps that satisfy \eebf\ are harmonic in their 
own induced metric, i.e.\ they satisfy the harmonicity condition with respect 
to the induced connection 
$$\christ cab=h^{cd}\p_dx^\mu\,g_{\mu\nu}\left(\p_a\p_bx^\nu+\christ\nu\sigma
\rho\p_ax^\sigma\p_bx^\rho\right).\eqn\eebh$$
The theory is a non-minimal coupling of the harmonic topological sigma model 
to topological gravity.   Note also that our gauge-fixing condition coincides 
with the equation of motion of the bosonic Nambu-Goto string.  Once again, we 
are constructing a theory whose path integral will be localized to the moduli 
spaces of all classical solutions of an associated bosonic theory, in this 
case the Nambu-Goto string theory.  

The gauge-fixing condition \eebf\ has $D-2$ independent components, as it 
should, since two components of $x^\mu$ should stay unfixed by virtue of 
worldsheet diffeomorphism invariance.  There are two constraints on 
$\Delta x^\mu$, expressing the fact that $\Delta x^\mu$ (as the trace of the 
second fundamental form of $\Phi$) is normal to $\Phi(\Sigma)$: 
$$\p_ax\cdot\Delta x=0.\eqn\eebi$$
Our gauge fixing condition is a section of the normal bundle $\CN$ to the 
worldsheet.  Since we will be encountering sections of $\CN$ very 
frequently, we introduce a special notation for the inner product induced in 
$\CN$ by the target metric $g_{\mu\nu}$; from now on, we write 
$$v\ast w\equiv v\cdot w-v\cdot\p_ax\,h^{ab}\,\p_bx\cdot w\eqn\eebj$$
for the inner product of the normal parts of any two vectors $v^\mu$ and 
$w^\mu$ from $\Phi^{-1}(TM)$.  

The gauge fixing condition requires us to introduce antighosts and 
auxiliaries,
$$\{Q,\chi^\mu\}=B^\mu,\qquad[Q,B^\mu]=0.\eqn\eebk$$
Since the gauge fixing function $\Delta x^\mu$ is a section of $\CN$, so are 
$\chi^\mu$ and $B^\mu$.  The fermionic gauge symmetry \eebd\ extends to the 
antighost multiplet by
$$\delta_\epsilon\chi^\mu=0,\qquad \delta_\epsilon B^\mu=\epsilon^a\p_a
\chi^\mu.\eqn\eebl$$ 

\section{Restoration of the Ghost-Antighost Symmetry}

Unlike harmonic topological sigma models, topological rigid string theory 
does not exhibit manifest symmetry between its ghosts and antighosts.  The 
ghost field $\psi^\mu$ represents infinitesimal deformations of the map 
$\Phi$, and is a section of $\Phi^{-1}(TM)$.  The antighost field, as we have 
just seen, is a section of the same bundle as the gauge fixing function 
$\Delta x^\mu$.  In harmonic topological sigma models, $\Delta x^\mu$ was 
also a section of $\Phi^{-1}(TM)$.  In the topological rigid string, 
worldsheet diffeomorphism invariance makes $\Delta x^\mu$ a section of the 
normal bundle $\CN\subset\Phi^{-1}(TM)$, hence spoiling the symmetry between 
ghosts and antighosts. 

There are several motivations for restoration of the ghost-antighost 
symmetry:  

\item{1.} The harmonic topological sigma model can be naturally interpreted 
as a deformation of the double-topological holomorphic sigma model, and as a 
twisted $\CN=4$ supersymmetric theory.  Similar structure can be expected in 
the topological rigid string.  In the double-topological theory, both the 
ghosts and the antighosts are members of the same BRST multiplet, and must 
be sections of the same bundle.  Also, the theory is then described in terms 
of a single BRST multiplet, in analogy with a similar property of the 
Yang-Mills theory (cf.\ \S{1.2}).  

\item{2.} The normal bundle $\CN$ is not always well-defined.  In particular, 
generic maps to two-dimensional targets are not immersions, which makes $\CN$ 
always ill-defined.  In explicit calculations, it is more convenient to deal 
with sections of the regular bundle $\Phi^{-1}(TM)$ instead.  

\item{3.} The existence of two BRST charges $Q,\bar Q$ in the harmonic 
topological sigma model have allowed us to write its Lagrangian as a very 
simple double commutator.  A similar formula will hold for the Lagrangian 
of the topological rigid string, and will allow us to draw interesting 
analogies between the topological rigid string theory and the physics of 
polymers, disordered systems and stochastic quantization.  

\item{4.} In the formulation with manifest symmetry between ghosts and 
antighosts, the overall ghost number anomaly is manifestly zero, which 
leads to a simple selection rule on physical correlation functions, and 
effectively decouples observables with positive ghost number from observables 
with ghost number zero (such as string winding modes).  

The worldsheet gauge invariance that spoils the symmetry between ghosts 
and antighosts comes to the rescue, and allows us to restore this symmetry.  
Since the longitudinal part of $\psi^\mu$ is a pure gauge of the ghostly 
gauge symmetry, the gauge-invariant parts of $\psi^\mu$ and $\chi^\mu$ are 
both sections of $\CN$.  Instead of gauge fixing the longitudinal part of 
$\psi^\mu$, we can go in the opposite direction and restore the symmetry 
between ghosts and antighosts by enlarging the gauge symmetry.  With this in 
mind, define 
$$\eqalign{\delta_\varepsilon x^\mu&=0,\cr
\delta_\varepsilon\chi^\mu&=\varepsilon^a\p_ax^\mu,\cr}\qquad
\eqalign{\delta_\varepsilon\psi^\mu&=0,\cr
\delta_\varepsilon B^\mu&=\varepsilon^a\p_a\psi^\mu.\cr}\eqn\eebm$$
This symmetry is imposed as yet another local fermionic symmetry of the 
theory.  Instead of being set to zero from the outset, the longitudinal part 
of $\chi^\mu$ is now a pure gauge of the new gauge symmetry.  

Just like in harmonic topological sigma models, the restoration of the 
symmetry between ghosts and antighosts allows us to define a second fermionic 
nilpotent charge $\bar Q$:
$$\eqalign{[\bar Q,x^\mu]&=\chi^\mu,\cr\{\bar Q,\psi^\mu\}&=-B^\mu,\cr}\qquad
\eqalign{\{\bar Q,\chi^\mu\}&=0,\cr[\bar Q,B^\mu]&=0.\cr}\eqn\eebn$$
Together with $Q$ they form an extended supersymmetry algebra,
$$\{Q,\bar Q\}=0,\qquad Q^2=0,\qquad\bar Q^2=0.\eqn\eebo$$
All fields of the model fall into an irreducible representation of \eebo .  

\section{Itoi-Kubota Symmetry}

In the previous section we extended the gauge symmetry of the model, in order
to restore the symmetry between ghosts and antighosts.  In this process we 
have actually obtained more than we required.  In addition to the new 
fermionic gauge symmetry, we have introduced a new bosonic gauge symmetry, 
$$\delta_vB^\mu=v^a\p_a x^\mu,\qquad\delta_v({\rm other\ fields})=0.\eqn\eebp
$$
This new symmetry is produced from the two fermionic gauge symmetries \eebm\ 
and \eebd, \eebl\ as their anticommutator, 
$$\delta_v\propto\{\delta_\epsilon,\delta_{\varepsilon}\}.\eqn\eebq$$
It is indeed a local symmetry, and allows us to consider $B^\mu$ a section 
of $\Phi^{-1}(TM)$, by making the longitudinal part of $B^\mu$ a gauge 
artifact.  

It is interesting to note that the bosonic gauge symmetry \eebp\ has actually 
been introduced in the bosonic rigid string theory quite some time ago, by 
Itoi and Kubota [\itoikub].  The original motivation of the authors of 
[\itoikub] for introducing this gauge symmetry was quite different from ours, 
however.  Here we have seen how the Itoi-Kubota symmetry naturally emerges in 
the supersymmetry algebra of the topological rigid string, as a consequence 
of the symmetry between ghosts and antighosts.  

\section{The Theory}

The topological rigid string Lagrangian can be written as a sum of two 
parts.  The first part is given by 
$$\eqalign{\CL_1&=\frac{1}{\alpha_0}\ints\rth\left\{h^{ab}\nabla_aB\cdot\p_bx
-h^{ab}\nabla_a\chi\cdot\nabla_b\psi-R_{\mu\sigma\rho\nu}\;h^{ab}\p_ax^\sigma
\p_bx^\rho\;\chi^\mu\psi^\nu\right.\cr
&\qquad\left.{}+\left(h^{ab}h^{cd}-h^{ac}h^{bd}-h^{ad}h^{bc}\right)\nabla_a
\psi\cdot\p_bx\;\nabla_c\chi\cdot\p_dx+\Delta x^\mu g_{\mu\nu}\christ\nu\sigma
\rho\chi^\sigma\psi^\rho\right\}.\cr}\eqn\eebr$$
$\CL_1$ is linear in the auxiliary field $B^\mu$ and quadratic in the 
fermionic fields $\psi^\mu$ and $\chi^\mu$, and all these fields can in 
principle be integrated out.  The integral over $B^\mu$ gives a delta 
function localized to the moduli space of minimal-area maps, while the 
integral over the fermions produces a volume element on the moduli space.  
$\CL_1$ is of course constructed as an exact BRST commutator, 
$$\CL_1=\{Q,\Psi_1\}=\{Q,\frac{1}{\alpha_0}\ints\rth h^{ab}\nabla_a\chi\cdot
\p_bx\}.\eqn\eebs$$

The second part of the topological rigid string Lagrangian smears out the 
delta fuction by introducing a term that is essentially $\propto B^2$, made 
covariant under all local symmetries.  Its full expression in terms of all 
fields is quite complicated, and we only write it here implicitly as a BRST 
commutator, 
$$\CL_2=\{Q,\Psi_2\},\eqn\eebt$$
with $\Psi_2$ given by
$$\eqalign{\Psi_2&=\frac{1}{\alpha_0}\ints\rth\left\{\chi\ast\left(B-
\christ{}\sigma\rho\chi^\sigma\psi^\rho\right)\right.\cr
&\qquad\left.{}+h^{ab}\left(\psi\ast\chi\;\nabla_a\chi\cdot\p_bx+\psi\ast
\nabla_a\chi\;\chi\cdot\p_bx-\chi\ast\nabla_a\chi\;\psi\cdot\p_bx\right)
\right\}.\cr}\eqn\eebu$$
One can find the full form of the Lagrangian by performing the BRST 
commutator explicitly, if one wishes so.  

The expression for $\CL$ becomes manageable when we keep only the 
diffeomorphism symmetry, and fix all the other gauge symmetries in a special 
gauge.  A particularly natural gauge choice for both of the fermionic gauge 
symmetries and the Itoi-Kubota symmetry is
$$\psi\cdot\p_ax=0,\qquad\chi\cdot\p_ax=0,\qquad B\cdot\p_ax=0.\eqn\eebua$$
In other words, we have used the gauge symmetries to set the longitudinal 
components of all worldsheet fields to zero.  In this particular gauge, the 
Lagrangian simplifies and can be explicitely written as follows:
$$\eqalign{\CL'&=\CL_1+a\CL_2=\frac{1}{\alpha_0}\ints\rth\left\{\left(
\vphantom{\frac{}{}}-B\cdot\Delta x-h^{ab}\nabla_a\chi\cdot\nabla_b\psi
-R_{\mu\sigma\rho\nu}h^{ab}\p_ax^\sigma\p_bx^\rho\,\chi^\mu\psi^\nu\right.
\right.\cr\vphantom{\int}&\left.{}+(h^{ab}h^{cd}-h^{ac}h^{bd}-h^{ad}h^{bc})
\nabla_a\psi\cdot\p_bx\;\nabla_c\chi\cdot\p_d x+\Delta x^\mu g_{\mu\nu}
\christ\nu\sigma\rho\chi^\sigma\psi^\rho\right)(1+a\psi\cdot\chi)\cr
\vphantom{\int}&{}+a\left(\vphantom{\frac{}{}}-B^2+2B^\mu g_{\mu\nu}\christ
\nu\sigma\rho\chi^\sigma\psi^\rho-R_{\mu\nu\sigma\rho}\chi^\mu\psi^\nu
\chi^\sigma\psi^\rho-\chi^\sigma\psi^\rho\christ{}\sigma\rho\ast\christ{}\mu
\nu\chi^\mu\psi^\nu\right.\cr
\vphantom{\int}&\qquad\qquad\qquad{}+(\psi\cdot\nabla_a\chi)\,h^{ab}\,
(\nabla_b\psi\cdot\chi)-(\psi\cdot\nabla_a\psi)\,h^{ab}\,(\nabla_b\chi\cdot
\chi)\cr
\vphantom{\int}&{}+h^{ab}\nabla_a\psi\cdot\p_bx\,(-\chi\cdot B+\chi^\mu 
g_{\mu\nu}\christ\nu\sigma\rho)+h^{ab}\nabla_a\chi\cdot\p_bx\,(\psi\cdot B
-\psi^\mu g_{\mu\nu}\christ\nu\sigma\rho\chi^\sigma\psi^\rho)\cr
\vphantom{\int}&\qquad\qquad\qquad\left.\left.{}+\vphantom{\frac{}{}}(\psi
\cdot\nabla_a\chi-\chi\cdot\nabla_a\psi)\,h^{ab}\,\p_bx^\mu g_{\mu\nu}
\christ\nu\sigma\rho\psi^\sigma\chi^\rho\right)\right\}.\cr}\eqn\eebub$$
$B^\mu$ can of course be integrated out, its equation of motion being 
$$\eqalign{\vphantom{\int}B^\mu&=-\frac{1}{2a}(1+a\psi\chi)\Delta x^\mu+
\frac{1}{2}\chi^\mu(\nabla\psi\cdot\p x)-\frac{1}{2}\psi^\mu(\nabla\chi\cdot
\p x)\cr
&\qquad\qquad\qquad\qquad{}+\left(\delta^\mu_\nu-\p_ax^\mu h^{ab}\p_b 
x^\lambda g_{\lambda\nu}\right)\christ\nu\sigma\rho\chi^\sigma\psi^\rho.\cr}
\eqn\eebuc$$
The explicit expression \eebub\ for the Lagrangian is not very illuminating, 
and is presented here only for completeness.  For all practical puproses, the 
only important properties of the Lagrangian are: 

\item{1.} The theory has been constructed as a topological string theory, 
according to the prescription of the Mathai-Quillen formalism.  In this 
sense, the model is (formally) exactly integrable since its partition 
functions calculate an equivariant Euler number of the moduli spaces 
[\mathaiq-\blau].  

\item{2.} Even without invoking the topological character of the theory, we 
will be able to rewrite the Lagrangian in a surprisingly simple form, 
amenable to a simple physical interpretation (see eqns.~ (4.29) -- (4.31) 
below).  

\item{3.} The theory is a topological version of the rigid string theory.  

Later on, it will prove useful to set $a=1$ and add another BRST exact term 
to the Lagrangian.  After that, the full Lagrangian is given by 
$$\eqalign{\CL=\{Q,&\frac{1}{\alpha_0}\ints\frac{\rth}{(1-\psi\ast\chi)^2}
\left\{h^{ab}\nabla_a\chi\cdot\p_bx+\chi\ast\left(B-\christ{}\sigma\rho
\chi^\sigma\psi^\rho\right)\right.\cr
&\qquad\left.{}+h^{ab}\left(\psi\ast\nabla_a\chi\;\chi\cdot\p_bx-\chi\ast
\nabla_a\chi\;\psi\cdot\p_bx-\psi\ast\chi\;\nabla_a\chi\cdot\p_bx\right)
\right\}.\cr}\eqn\eebv$$
The new term that has been added to the original Lagrangian contains only 
higher-order terms in the fermi fields, and does not bring in any new 
branches of the moduli spaces from the infinity in the space of all field 
configurations.  Hence, the term has been designed in such a way that its 
addition to the Lagrangian should not change the value of the path integral.  

It is instructive to look at the bosonic sector of the resulting Lagrangian.  
Upon setting all fermi fields to zero, both $\CL$ and $\CL'$ simplify to 
$$\CL_0=\frac{1}{\alpha_0}\ints\rth\left(-B\cdot\Delta x-B\ast B\right).
\eqn\eebw$$
Integrating out the auxiliary fields $B^\mu$, one gets 
$$\CL_0=\frac{1}{4\alpha_0}\ints\rth\Delta x\cdot\Delta x,\eqn\eebx$$
which is the Lagrangian of the bosonic rigid string theory (at zero string 
tension).  Hence, our topological theory, derived from the requirement of 
localization to the moduli spaces of minimal-area maps, can indeed be 
considered a topological version of the rigid string theory.  Remarkably, the 
bosonic rigid string theory (in four target dimensions) has been studied some 
time ago by Polyakov [\polya], Kleinert [\kleinert] and others [\otherrs], as 
a candidate for QCD string theory.  

Although its explicit component form is complicated, the full Lagrangian of 
the topological string theory can be written in an extremely simple and 
form, using the second supersymmetry charge $\bar Q$ of \eebn .  This 
supercharge allows us to write the Lagrangian as a double commutator.  
Thus, we can write $\CL_1$ of \eebr\ as
$$\CL_1=\{Q,[\bar Q,\frac{1}{\alpha_0}\ints\rth]\}.\eqn\eeby$$
Similarly, the second part of the Lagrangian, as given by \eebt\ and \eebu , 
can be written as 
$$\CL_2=\{Q,[\bar Q,\frac{1}{\alpha_0}\ints\rth\chi\ast\psi]\},\eqn\eebz$$
and the deformed Lagrangian \eebv\ takes the form
$$\CL=\{Q,[\bar Q,\frac{1}{\alpha_0}\ints\frac{\rth}{1-\psi\ast\chi}]\}.
\eqn\eebzz$$
Thus, the whole Lagrangian of the topological rigid string can be written as 
a simple double commutator, and its invariance under the $\u 1$ symmetry that 
mixes ghosts and antighosts is now manifest.  The simple form of \eeby\ -- 
\eebzz\ is the key to an analogy between the topological rigid string theory 
and some models studied in the physics of polymers and disordered systems 
[\future], and leads to a quite non-trivial physics in higher target 
dimensions.  

\section{Partition Functions and Correlation Functions}

The partition function of the topological rigid string theory for a given 
worldsheet $\Sigma$ and target $M$ calculates the equivariant Euler number of 
the (regular locus of the) moduli spaces of minimal-area maps from $\Sigma$ 
to $M$.  Similarly as in the harmonic topological sigma models, this claim 
can be confirmed by a direct semiclassical calculation at $\alpha_0\ll 1$.  
A more elegant way to prove it is to notice that the theory has been 
constructed as an infinite-dimensional version of the Mathai-Quillen 
formalism, which is essentially a specific algorithm how to calculate Euler 
characters of vector bundles over manifolds (and their equivariant analogs, 
if there is a symmetry group acting on the vector bundle).  The fact that 
topological field theories are infinite-dimensional versions of the 
Mathai-Quillen formalism has been first discussed by Atiyah and Jeffrey in 
[\atiyahj].  Since this interpretation of topological field theories is well 
covered in the literature and is now considered standard, it will not be 
discussed here. (For an excellent short review aimed at physicists, see 
[\blau]; more details can be found in [\mathaiq-\ezra].)  

Consider a minimal-area map $\Phi$, given in coordinates by $x^\mu$. Any 
deformation $x^\mu{}'=x^\mu+\delta x^\mu$ that is still a minimal area map 
to lowest order in $\delta x^\mu$ must satisfy the linearized minimal-area 
equation 
$$\CJ^\mu_\nu\;\delta x^\nu=0,\eqn\eebwwa$$
where
$$\CJ^\mu_\nu \equiv\left\{\left(\delta^\mu_\nu-\p_cx^\mu h^{cd}\p_dx^\lambda 
g_{\lambda\nu}\right)\Delta-\left(g^{\mu\lambda}-\p_cx^\mu h^{cd}\p_d
x^\lambda\right)R_{\lambda\sigma\rho\nu}\;h^{ab}\p_ax^\sigma\p_bx^\rho
\right\}.\eqn\eebww$$
(Here we have chosen the gauge $\delta x\cdot\p_ax=0$, in order to eliminate 
the apparent deformations of $\Phi$ that correspond to reparametrizations of 
$\Sigma$.)  This equation happens to be identical to the equation that 
defines the fermionic zero modes, i.e.\ it coincides with the linearized 
equation of motion for both the ghost and antighost fields (in the gauge 
\eebua ).  Even in the theory of minimal-area maps, this equation is called 
the Jacobi equation, and its solutions are tangent to the moduli spaces of 
minimal-area maps.  

Because of the symmetry between ghosts and antighosts, there is always an 
equal number of ghost and antighost zero modes.  Hence, as in the sigma model 
case, there is no ghost number anomaly, and observables with positive ghost 
numbers effectively decouple from observables with ghost number zero.  

\chapter{Topological Rigid Strings in Two Dimensions}

In this section, we continue our discussion of the topological rigid string, 
in the restricted class of two dimensional targets.  In the next section, the 
partition functions of the topological rigid string theory will be compared 
to the results of the large $N$ expansion in 2D QCD.  

\section{Suppression of Folds and Worldsheet Self-Avoidance}

The functional integral of the topological rigid string theory is localized 
to the infinitesimal vicinity of the moduli spaces of all solutions to the 
gauge-fixing constraint, which in our case is 
$$\Delta x^\mu=0.\eqn\eeca$$
In $D$ target dimensions, this condition has just $D-2$ independent 
components, as a result of worldsheet diffeomorphism invariance.  In two 
target dimensions, we are apparently left with $D-2=0$ conditions!  Yet, the 
condition \eeca\ is non-trivial even in two dimensions.  The naive counting 
of independent components of \eeca\ was based on the transversality condition 
$$\p_ax\cdot\Delta x\equiv 0,\eqn\eecaca$$
which represents two constrains on $\Delta x$ whenever $\p_ax^\mu$ is 
non-degenerate as a two-by-two matrix (i.e.\ in those points where $\Phi$ is 
an immersion).  In two dimensions, generic maps are not immersions, and their 
induced metric is always degenerate somewhere.  This fact makes the condition 
\eeca\ non-trivial even in two dimensions.  \eeca\ is the equation of motion 
of the Nambu-Goto string and represents the minimal-area (or more precisely, 
critical-area) condition on $\Phi$.  Maps with folds of non-zero length are 
not critical-area maps, hence they violate \eeca\ and do not contribute to 
the path integral.  Thus, in two dimensions, the sole purpose of the 
minimal-area condition \eeca\ is to suppress maps with folds of non-zero 
length.%
\foot{In general, the minimal-area equation \eeca\ requires interpretation, 
and a mathematically precise specification of the class of maps that are 
considered solutions of \eeca\ is a subtle issue [\fomenko].  It leads to a 
necessary extension of the naive definition of a surface to more general 
objects, such as stratified surfaces and multivarifolds [\fomenko].  In this 
paper we will try to avoid introducing these complicated mathematical 
objects, and only note here that our definition of minimal-area maps is that 
of stratified surfaces.}  

The suppression of folds has been identified as one of the crucial properties 
of the large-$N$ expansion of 2D QCD, and is on general grounds (such as the 
strong coupling expansion on the lattice) expected from higher dimensional 
QCD string theory as well.  The absence of folds in string theory means that 
the string worldsheet is self-avoiding.  This self-avoidance is local on the 
worldsheet; in particular, it is different from spacetime self-avoidance of 
the strings, since two distinct worldsheet points are still allowed to occupy 
the same spacetime point.  As it turns out, the topological rigid string can 
be derived from the bosonic Nambu-Goto string by imposing a simple condition 
of worldsheet self-avoidance on the latter [\future].  

\section{Moduli Spaces of Minimal-Area Maps}

We have treated diffeomorphism invariance of the theory as an equivariant 
symmetry, therefore we must keep it a manifest symmetry of our theory.  In 
particular, the moduli spaces must be parametrized in a diffeomorphism 
invariant way.   Fortunately, this requirement is a virtue rather than a 
constraint, and will allow us to understand the geometry of the moduli spaces 
in detail.  (Here we follow the insight of Gross and Taylor, who parametrized 
the moduli spaces emerging from the large $N$ expansion of 2D QCD in a 
closely related way.)  With a simple parametrization of the moduli spaces at 
hand, we will be able to calculate their Euler numbers.  

Since maps with no folds of non-zero length are coverings of $M_G$ 
almost everywhere, we obtain a very simple classification of maps that 
solve \eeca .  In a given homotopy class $[\Sigma_g,M_G]$, minimal-area 
maps are coverings of $M_G$ outside a fininte number of points $P_1,\ldots,
P_k\in M_G$.  At each of these points, a generic map exhibits one of the 
following moduli:

\item{1.} A simple branchpoint of degree one.  (We define the degree of a 
branchpoint as the number of covering sheets above a generic target point 
minus the number of sheets above the branchpoint.)  

\item{2.} A collapsed handle.  (Whenever a handle of $\Sigma_g$ is mapped 
to $M_G$ in a homotopically trivial way, the minimal-area condition (in the 
sense of stratified surfaces [\fomenko]) requires the handle to be mapped to 
a point in $M_G$.) 

\item{3.} A collapsed disk.  (A map with a homotopically trivial domain of 
$\Sigma_g$ mapped into one point; although these maps also satisfy the 
minimal-area condition, we will see below that their contribution to the 
partition function is in fact zero.)  

In addition to maps with various combinations of these moduli, other 
solutions of $\Delta x^\mu=0$ exist, and are actually quite important in the 
path integral.  These additional solutions represent maps of ``critical 
area'' rather than minimal area, and are ustable solutions of the associated 
bosonic theory (although they are of course absolute minima of the 
topological rigid string action).  The fact that unstable solutions of the 
bosonic theory will contribute to the partition function is yet another 
similarity to the Yang-Mills theory in spacetime.  

The critical-area maps add one more modulus type to the ones listed above: 

\item{4.} A collapsed neck between two sheets of opposite orientation.  

Henceforth we call the moduli of these four types the ``simple moduli,'' to 
distinguish them from the moduli that emerge when two or more simple moduli 
coalesce (i.e.\ at the compactification locus of the moduli spaces).  Notice 
also that maps from $\Sigma$ to $M$ with any combination of the simple moduli 
are all smooth maps.%
\foot{While the differential topology of the maps is smooth, the induced 
metric is of course singular.}

The moduli spaces of critical-area maps can then be described as follows.  
Consider a map that covers $M_G$ by $n$ ($\tilde n$) sheets of the same 
(opposite) orientation everywhere except in $k$ distinct points 
$(P_1,\ldots,P_k)$, and pick one simple modulus for each $P_i$.  To fix the 
homotopy class uniquely, in addition to the simple moduli at $P_i$ we must 
specify how the sheets of the cover are permuted when we go around any 
non-contractible loop in $M_G$.  Hence, we choose $4G$ elements $a_j,b_j$, 
$j=1,\ldots,2G$ of the group of permutations of the cover, 
$S_n\otimes S_{\tilde n}$.  Altogether, these choices are constrained by one 
homotopy condition, 
$$\prod_{i=1}^kp_i(P_i)\prod_{i=1}^Ga_jb_ja_j^{-1}b_j^{-1}=1\quad\in
\quad S_n\otimes S_{\tilde n},\eqn\eepp$$
where $p_i(P_i)$ is the element of $S_n\otimes S_{\tilde n}$ that represents 
the permutation of the covering sheets when we go around $P_i$.  When the 
simple modulus at $P_i$ is a simple branchpoint, $p_i(P_i)$ is a permutation 
of two sheets of identical orientation; permutations associated with the 
other three types of simple moduli are trivial. 

Each of the moduli is point-like in the target, and its target location $P_i$ 
serves as a natural coordinate in the bulk of the moduli space.  For a given 
choice of data that satisfy \eepp , define 
$$\tilde\CM^0_{G,k}=\left\{(P_1,\ldots,P_k),\quad P_i\in M_G,\quad P_i\neq 
P_j\ {\rm for}\ i\neq j\right\}.\eqn\eecc$$
As we will see later, the removed points provide a natural compactification 
of $\tilde\CM^0_{G,k}$.  

The actual bulk moduli spaces $\CM^0_{G,k}$ are factors of 
$\tilde\CM^0_{G,k}$ by the permutation group $S_k$ that permutes $P_i$'s, 
$$\CM^0_{G,k}=\tilde\CM^0_{G,k}/S_k.\eqn\eecb$$
From the point of view of the worldsheet path integral, this factorization 
has two different sources: 

\item{1.} When two simple moduli of the same type are on the same connected 
component of the worldsheet, they are indistinguishable (i.e.\ they are 
related by a global worldsheet diffeomorphism).  

\item{2.} Two different simple moduli on the same connected worldsheet 
components, as well as two simple moduli of the same type but on different 
non-isomorphic worldsheet components, are distinguishable.  Their 
contribution is however overcounted when we sum independently over all 
possible simple moduli at all $P_i$'s.  In order to avoid this overcounting, 
we must factorize $\tilde\CM^0_{G,k}$ by $S_k$.  

The non-compact bulk moduli spaces $\CM^0_{G,k}$ (as well as their finite 
coverings $\tilde\CM^0_{G,k}$) have a natural compactification dictated by 
the theory itself.  Whenever two or more simple moduli coalesce in one point 
$P$ in $M_G$, a composite modulus is formed.  Among the composite moduli are 
maps with several branchpoints of degree one at a given location $P$, maps 
with a branchpoint of higher degree at $P$, maps with collapsed manifold of 
higher genus at $P$, maps with twisted connecting necks between sheets of 
different orientation, etc.; the structure of all possible composite moduli 
is uniquely determined by the simple moduli.  The locations of composite 
moduli serve as coordinates on the compactification locus of the moduli 
spaces $\CM^0_{G,k}$.  Thus, given a component of the bulk moduli space 
$\CM^0_{G,k}$, all possible composite moduli compactify the cover 
$\tilde\CM^0_{G,k}$ to $\tilde\CM_{G,k}$, 
$$\tilde\CM_{G,k}\equiv(M_G)^k=\left\{(P_1,\ldots,P_k),\quad P_i\in M_G
\right\}.\eqn\eeccw$$
The factorization of $\tilde\CM_{G,k}$ by the symmetry group that permutes 
the copies of $M_G$ turns this component of the compactified moduli space 
into an orbifold:
$$\CM_{G,k}=\tilde\CM_{G,k}/S_k.\eqn\eeccz$$

So far we have compactified a single bulk component of the moduli space, with 
the homotopy class of $\Phi$ fixed uniquely by our fixed choice of $a_j,b_j$ 
and the simple moduli at $P_i$.  Since in general a given composite modulus 
can be created in several ways when different groups of simple moduli 
coalesce, two or more different bulk components of the total moduli space can 
have a common compactification locus.  Consequently, the total moduli space 
is strictly speaking not even an orbifold, but it can always be decomposed 
into orbifolds, and its Euler number can always be uniquely defined using 
this decomposition.  

Several other facts are worth noticing:  

\item{1.} The parametrization of the moduli spaces by the target location 
of the allowed singularities is manifestly invariant under worldsheet 
diffeomorphisms.  

\item{2.} Since this parametrization of the moduli spaces does not allow us 
to keep track of the connectivity of the worldsheet, our partition function 
sums over all worldsheet topologies, not only the connected ones.  This and 
the connected partition function are of course related to each other 
exponentially.  

\section{Euler Numbers of the Moduli Spaces}

The partition functions of the topological rigid string theory calculate the 
Euler number of the moduli spaces.  As we have seen, the structure of the 
moduli spaces is in fact quite simple, and we can calculate their Euler 
numbers directly, for example by cell decomposition.  The calculation 
presented here is very similar to the analogous calculation done by Cordes, 
Moore and Ramgoolam for Hurwitz moduli spaces in [\cmr].  

First, let us recall the definition of the Euler number of an orbifold 
[\thurston].  Just as manifolds are locally modelled by regions $\CU_i$ in 
${\bf R}^n$, 
an orbifold $\CO$ is locally modelled by regions $\CU_i$ of ${\bf R}^n/G_i$, 
where $i$ goes over the set of all coordinate systems on $\CO$, and $G_i$ 
is a finite group that acts on ${\bf R}^n$.  Coordinate changes are required 
to respect the group action by $G_i$ in a natural manner, which allows us 
to define for each point $x$ in $\CO$ a group $G_x$, called the ``isotropy 
group'' of $x$, as the smallest $G_i$ associated with a domain containing  
$x$.  With this notation, the Euler number of $\CO$ is defined as follows.  
Pick a cell decomposition of $\CO$ which respects the isotropy groups on 
$\CO$, i.e.\ all points in a given cell $\CC$ have the same isotropy group, 
which we denote by $G_\CC$.  The Euler number is then given by a sum over all 
cells, 
$$\chi(\CO)=\sum_\CC(-1)^{{\rm dim}\,\CC}\frac{1}{|G_\CC|}.\eqn\eecgga$$
This definition of the Euler number is natural with respect to products and 
disjoint unions of orbifolds, a fact that will be used below.  

Using \eecgga , the orbifold Euler number of $\modspace k$ can then be 
calculated as follows: 
$$\chi(\modspace k)=\frac{1}{k!}\left\{\chi(\tilde\CM^0_{G,k})\right\}+\ldots
=\frac{1}{k!}(2-2G)^k.\eqn\eecg$$
The expression in the parentheses is the Euler number of the locus in 
$(M_G)^k$ on which $S_k$ acts freely (``free locus'' from now on), while the 
dots represent contributions from the subset of $(M_G)^k$ where at least two 
$P_i$'s coincide, i.e.\ from the points with non-zero isotropy group.  

The Euler number of $\CM^0_{G,k}$ can be easily calculated by induction in 
$k$.  For $k=2$, the Euler number is easily computed directly.  First we use 
the multiplicativity property of the Euler number to 
get $\chi((M_G)^2)=[\chi(M_G)]^2=(2-2G)^2$, and then, using the additivity of 
$\chi$, we subtract from this result the Euler number of the diagonal part of 
$(M_G)^2$, which is equal to $2-2G$.  That gives the Euler number of the free 
locus in $(M_G)^2$.  Since $S_2$ acts on the free locus freely, the Euler 
number of $\CM^0_{G,2}$ (which is the factor of the free locus by $S_2$) is 
$1/(2!)$ times the Euler number of $\tilde\CM^0_{G,k}$: 
$$\chi(\CM^0_{G,2})=\frac{1}{2}\left\{(2-2G)^2-(2-2G)\right\}=\frac{(2G-2)
(2G-1)}{2}.\eqn\eech$$
For $k=3$, the direct calculation is still simple, and gives 
$$\eqalign{\chi(\CM^0_{G,3})&=\frac{1}{3!}\left\{(2-2G)^3-3\left[(2-2G)^2-
(2-2G)\right]-(2-2G)\right\}\cr
&\qquad{}=-\frac{(2G-2)(2G-1)2G}{6}.\cr}\eqn\eeci$$
In this expression, $(2-2G)^3$ is the Euler number of $(M_G)^3$, the 
term in the brackets subtracts the Euler number of the three subsets in 
$(M_G)^3$ where exactly two $P_i$'s coincide as elements of $M_G$, while 
the last term subtracts the Euler number of the diagonal $M_G$, which is the 
set of points where all three coordinates $P_1,P_2,P_3$ coincide as elements 
of $M_G$.  

In order to derive the general formula, assume first that we have calculated 
the Euler number of $\CM^0_{G,k}$; the Euler number of $\CM^0_{G,k+1}$ is 
then calculated as follows.  We can represent $\tilde\CM^0_{G,k+1}$ as 
$M_G\times\tilde\CM^0_{G,k}$, minus the set of diagonal points.  There are 
exactly $k$ possibilities how the added point can coincide with another point 
as an element of $M_G$, and each possibility leads to a subspace of $M_G\times
\tilde\CM^0_{G,k}$ isomorphic to $\tilde\CM^0_{G,k}$ itself.  Since these $k$ 
copies of $\tilde\CM^0_{G,k}$ are non-intersecting in $M_G\times\tilde
\CM^0_{G,k}$, we obtain the following recursion relation, 
$$\chi(\tilde\CM^0_{G,k+1})=\chi(M_G)\;\chi(\tilde\CM^0_{G,k})-k\,\chi(\tilde
\CM^0_{G,k}).\eqn\eecj$$
Since $\CM^0_{G,k}$ is a factor of $\tilde\CM^0_{G,k}$ by the free action of 
the permutation group $S_k$, the recursion relation \eecj\ can be rewritten 
as
$$(k+1)\chi(\CM^0_{G,k+1})=(2-2G-k)\chi(\CM^0_{G,k}).\eqn\eeck$$
This relation can be easily solved, and the general formula for the Euler 
numbers at arbitrary values of $G$ and $k$ finally is 
$$\chi (\CM^0_{G,k})=(-1)^k\pmatrix{2G+k-3\cr k\cr}.\eqn\eecl$$
We can summarize these Euler numbers in a generating formula, by 
introducing an auxiliary variable $x$ and defining $\chi(x)\equiv\sum\chi
(\CM^0_{G,k})\;x^k$.  Using the explicit expressions \eecl\ for the Euler 
numbers, we can write the generating function $\chi(x)$ in a surprisingly 
simple form: 
$$\chi(x)\equiv\sum_{k=0}^{\infty}\chi(\CM^0_{G,k})\;x^k=\sum_{k=0}^\infty
(-1)^k\pmatrix{2G+k-3\cr k\cr}\;x^k=\frac{1}{(1+x)^{2G-2}}.\eqn\eecm$$
This formula will prove very valuable in \S{6}, where we compare the 
partition functions of the topological rigid string theory with the results 
of the large-$N$ expansion in 2D QCD.  

\section{Partition Functions of the Topological Rigid String}

The partition function of the topological rigid string theory on a fixed 
target $M_G$ of genus $G$ contains a contribution from various components of 
the moduli spaces of minimal area maps as analyzed in the previous 
subsections.  We can impose the homotopy constraint \eepp\ in the form of a 
delta function, which allows us to write the partition function as an 
unrestricted sum over all possible moduli as well as values of the homotopies 
$a_j,b_j$, 
$$\CZ=\sum_{n,\tilde n}g_{\rm string}^{(n+\tilde n)(2G -2)}
\frac{1}{n!\tilde n!}\sum_{k=0}^\infty\sum_{a_j,b_j}\zeta_{G,k}\;
\delta\left(\sigma_{n,\tilde n}(P_1)\ldots\sigma_{n,\tilde n}(P_k)
\prod_{j=1}^Ga_jb_ja_j^{-1}b_j^{-1}\right).\eqn\eeswaq$$
Here $\sigma_{n,\tilde n}(P_i)$ is a sum over all possible moduli in $P_i$, 
each modulus being represented by its element of $S_n\otimes S_{\tilde n}$ 
and weighted by its contribution to the overall power of $g_{\rm string}$, 
and $\zeta_{G,k}$ are numbers that implicitly depend on all the data that are 
being summed over.  The intergration over the moduli spaces and over the 
integrable zero modes of the ghosts and antighosts in the topological rigid 
string theory gives the Euler number of the moduli spaces, and we expect 
$$\zeta_{G,k}\propto\chi(\CM^0_{G,k}).\eqn\eexpct$$
Before we write an explicit expression for $\sigma_{n,\tilde n}(P_i)$, let 
us analyze this expectation in detail.  Although essentially true, the naive 
statement \eexpct\ receives corrections from the integration over the 
remaining modes in the path integral.  

First we show that collapsed disks do not contribute to the partition 
function, since the corresponding moduli spaces have Euler number zero.  
Consider an arbitrary fixed configuration of moduli other than collapsed 
disks, in points $(P_1,\ldots,P_k)$ in the target.  Adding $s$ collapsed 
disks in additional points $(P'_{k+1},\ldots,P'_{k+s})$ does not change the 
genus of the worldsheet.  When two collapsed disks at $P'_{k+i},\ P'_{k+j}$ 
coalesce, they again form a collapsed disk.  In this sense, the moduli space 
of maps with $s$ collapsed disks is a compactification locus of the moduli 
space of maps with $s+1$ collapsed disks.  Hence, for fixed moduli at 
$(P_1,\ldots,P_k)$, there are two disconnected components of the moduli 
spaces:  one corresponds to maps with no collapsed disks, and consists of 
just one point for each set of moduli at fixed values of $(P_1,\ldots,P_k)$; 
the other corresponds to maps with an arbitrary number of collapsed disks.  
This second component of the moduli space is nominally infinitely 
dimensional, and all moduli spaces with finite $s$ are nested in it.  As 
$s\rightarrow\infty$, the Euler number of this moduli space goes to zero, and 
the only contribution to the partition function thus comes from maps with no 
collapsed disks.  

One reason why the Euler numbers of the moduli spaces are not the whole 
story comes from the existence of minimal-area maps with additional zero 
modes of the ghost-for-ghost $\phi^a$.  Path integrals in these sectors are 
notoriously hard to calculate since the zero modes of $\phi^a$ make the 
moduli space of all zero modes non-compact, but general arguments exist that 
these contributions typically vanish.  Even in the topological rigid string, 
this is a subtle issue, and its full clarification would go well beyond the 
scope of this paper.  To see which classes of maps lead to additional zero 
modes of $\phi^a$, we will use a BRST fixed point theorem.  

BRST fixed-point theorems (see e.g.\ [\wittenmir]) use the BRST invariance of 
the theory to argue that the only non-zero contribution to the path integral 
comes from infinitesimal vicinity of the set of configurations annihilated by 
the BRST charge (i.e.\ are ``fixed points'' of the BRST supersymmetry 
transformation).  Recall first the action of the equivariant BRST charge on 
the fields of the topological rigid string, 
$$[Q,x^\mu]=\psi^\mu,\qquad\{Q,\psi^\mu\}=\phi^a\p_ax^\mu.\eqn\eecn$$
Whenever a two-dimensional domain $\CD$ in $\Sigma$ is mapped to one point in 
the target (e.g.\ when the map has a collapsed handle or a collapsed disk), 
the two-by-two matrix $\p_ax^\mu$ is degenerate everywhere in $\CD$.  The 
condition that the BRST transformation of all fields be zero entails 
$$\psi^\mu_0=0,\qquad\phi^a_0\p_ax^\mu_0=0,\eqn\eeco$$
and a degenerate matrix $\p_ax^\mu_0$ leads to many non-trivial solutions for 
$\phi^a_0$.  

Additional zero modes of $\phi^a$ also exist for some maps with composite 
moduli, i.e.\ in some components of the compactification locus of the moduli 
spaces.  The simplest modulus leading to additional zero modes is generated 
when two identical simple branchpoints coalesce in $P$ and form a connecting 
tube between two sheets of the same orientation.  More complicated composite 
moduli of this type (i.e.\ twisted connecting tubes) are created when 
multiple branchpoints coalesce.  Maps with these composite moduli map a 
homotopically non-trivial loop on $\Sigma$ to the target point $P$.  Along 
this loop, the two-by-two matrix $\p_ax^\mu_0$ degenerates and non-trivial 
solutions $\phi^a_0$ of \eeco\ exist.  

Even on a regular component of the moduli spaces, the partition function 
is not necessarily equal to the Euler number, and can in fact differ from 
$\chi(\CM)$ by a sign.  Indeed, while for minimal-area maps the fermionic and 
bosonic one-loop determinants cancel each other exactly, an extra minus sign 
can appear for unstable critical-area map.  A generic unstable critical-area 
map contains at least one neck that connects two covering sheets of opposite 
orientations; such a map can be deformed to a map with lower area by opening 
the neck into a connecting tube of non-zero radius, with a fold of non-zero 
length.  This deformation corresponds to a negative eigenvalue of the 
fermionic operator \eebww , which modifies the contribution of these moduli 
spaces to 
$$(-1)^v\chi(\CM^0_{G,k}),\eqn\eeswaa$$
where $v$ is the number of simple connecting necks among the moduli at $P_1,
\ldots,P_k$.  

The contribution of the regular moduli spaces to the partition function can 
thus be summarized in the following expression for $\sigma_{n,\tilde n}(P)$, 
$$\eqalign{\sigma_{n,\tilde n}(P)&=\sum_{\sigma\otimes\tau\in S_n\otimes 
S_{\tilde n}}\sigma\otimes\tau\,g_{\rm string}^{n+\tilde n-K_\sigma-K_\tau}
\cr&\quad{}\times\prod_{\ell=1}^{{\rm min}\, (n,\tilde n)}\left(
\sum_{v_\ell=0}^{{\rm min}\,(\sigma_{(\ell)},\tau_{(\ell)})}(-1)^{v_\ell}
\ell^{v_\ell}v_\ell!\pmatrix{\sigma_{(\ell)}\cr v_\ell\cr}\pmatrix{\tau_{
(\ell)}\cr v_\ell\cr}g_{\rm string}^{2v_\ell}\right).\cr}\eqn\eessin$$
Here $\sigma_{(\ell)}$ and $\tau_{(\ell)}$ is the number of cycles of length 
$\ell$ in $\sigma$ and $\tau$, $K_\sigma$ is the number of all cycles in 
$\sigma$, and the sum is restricted to $\sigma$ and $\tau$ that are not 
simultaneously trivial.  

The combinatorial factors are present in \eessin\ to ensure that each 
component of the moduli spaces that contributes to \eessin\ does so exactly 
once.  More explicitly, the geometry of the factors is as follows.  $\ell$ 
denotes the number of sheets of the orientation-preserving and 
orientation-reversing cover that are connected by an orientation-reversing 
collapsed neck; this neck can be visualized as the simple 
orientation-reversing connecting neck, twisted $\ell$ times.  In the sum over 
all values of $\ell$, $v_\ell$ is the number of orientation-reversing 
collapsed necks that connect a cycle with $\ell$ sheets of a given 
orientation with a cycle of $\ell$ sheets with the opposite orientation, the 
binomial coefficients count the number of combinations in which the 
orientation-preserving and orientation-reversing cycles of length $\ell$ can 
be combined to form $v_\ell$ connecting necks, $v_\ell !$ represents all 
possible permutations of the combinations, and the additional power 
$\ell^{v_\ell}$ comes from the fact that a given cycle of length $\ell$ can 
be combined with a given cycle of the same length and opposite orientation 
in $\ell$ different ways.  The power of $g_{\rm string}$ in the sum over all 
$\ell$ just weighs the contribution of the necks to the overall Euler number 
of the worldsheet.  

Since in \eessin\ we have already absorbed into $\sigma_{n,\tilde n}(P)$ 
the additional minus signs $(-1)^v$ that come from the contribution of the 
negative modes to the one-loop determinant, each component of the moduli 
spaces that contributes in \eessin\ contributes exactly $\chi(\CM^0_{G,k})$.  
Hence, we can write our final formula for the partition functions of the 
topological rigid string as follows, 
$$\CZ=\sum_{n,\tilde n}\frac{1}{n!\tilde n!}g_{\rm string}^{(n+\tilde n)
(2G-2)}\sum_{k=0}^\infty\sum_{a_1,b_1\ldots a_G,b_G}\chi(\CM^0_{G,k})\;
\delta\left(\prod_{i=1}^k\sigma_{n,\tilde n}(P_i)\prod_{j=1}^Ga_jb_ja_j^{-1}
b_j^{-1}\right).\eqn\eess$$
This formula can be rewritten in terms of an integral over the total 
moduli space, 
$$\CZ=\int_\CM\e{-\CR_0(\CM)},\eqn\eezxzx$$
where $\CR_0(\CM)$ is a suitably defined two-form on the total moduli space 
$\CM$.  This two-form is the sum of the induced curvature two-forms along the 
directions that contribute to the partition functions, multiplied by an 
additional minus sign when the modulus is a connecting neck.  

\chapter{Large-$N$ QCD Strings in Two Dimensions}

We have seen in the previous sections how the topological rigid string 
theory is defined, and that its path integral gives the Euler numbers 
of moduli spaces of minimal-area maps.  Here we show how this reproduces the 
results of the large-$N$ expansion in 2D QCD.  

The large-$N$ expansion in the two-dimensional Yang-Mills theory with 
gauge group $\su N$ can be written as
$$\eqalign{\CZ(G,\lambda A,N)&=\sum_{n,\tilde n}\frac{1}{n!\tilde n!}
\e{-(n+\tilde n)\lambda A/2}\sum_{s,k}(-1)^s\frac{(\lambda A)^{s+k}}{s!k!}
N^{(n+\tilde n)(2-2G)-s-2k}\frac{(n-\tilde n)^{2k}}{2^k}\cr
&\qquad{}\times\sum_{p_1,\ldots p_s\in T_2}\sum_{a_1,b_1,\ldots a_G,b_G}
\delta(p_1\ldots p_s\,\Omega^{2-2G}_{n,\tilde n}\prod_{j=1}^Ga_jb_ja_j^{-1}
b_j^{-1})\cr}\eqn\eeda$$
Here the ``$\Omega$-points'' correspond to (in the notation of [\grta]) 
$$\Omega_{n,\tilde n}=1+\tilde\Omega_{n,\tilde n}=\sum_{\sigma,\tau}
\sigma\otimes\tau\sum_{v,v'}(-1)^{K_v}C_vN^{K_{\sigma\backslash v}+K_{\tau
\backslash v}-n-\tilde n}.\eqn\eewqwqa$$

\section{Zero Target Area/Zero String Tension}

At $\lambda=0$, the partition function can be written as a sum over the 
number of orientation-preserving and orientation reversing sheets $n$ and 
$\tilde n$, each contribution being equal to
$$\CZ_{n,\tilde n}(G,\lambda A=0,N)=\frac{1}{n!\tilde n!}N^{(n+\tilde n)
(2-2G)}\sum_{a_1,b_1,\ldots a_G,b_G}\delta(\Omega^{2-2G}_{n,\tilde n}
\prod_{j=1}^Ga_jb_ja_j^{-1}b_j^{-1})\eqn\eeqeae$$
Since $\Omega_{n,\tilde n}=1+\tilde\Omega_{n,\tilde n}$, we can expand 
$\Omega_{n,\tilde n}$ in the powers of $\tilde\Omega_{n,\tilde n}$, and write 
\eeqeae\ as 
$$\eqalign{\CZ_{n,\tilde n}(G,\lambda A&=0,N)=\frac{1}{n!\tilde n!}
N^{(n+\tilde n)(2-2G)}\cr
&\qquad{}\times\sum_{k=0}^{\infty}(-1)^k\pmatrix{2G+k-3\cr k\cr}
\sum_{a_1,b_1,\ldots a_G,b_G}\delta(\tilde\Omega_{n,\tilde n}^k
\prod_{j=1}^Ga_jb_ja_j^{-1}b_j^{-1})\cr}\eqn\eeqe$$
Recall the generating function of the Euler numbers of the moduli spaces of 
minimal area maps, \eecm .  Hence, the binomial factor in \eeqe\ is 
essentially the Euler number of the moduli space of maps that are covers of 
$M_G$ except in $k$ points, where $\tilde\Omega_{n,\tilde n}$ is inserted.  
The expression \eewqwqa\ for $\tilde\Omega_{n,\tilde n}$ is not very 
transparent, but luckily, it can be rewritten as [\grta] 
$$\Omega_{n,\tilde n}=\sum_{\sigma,\tau}\sigma\otimes\tau\,N^{-n-\tilde n
+K_\sigma+K_\tau}\prod_{\ell=1}^{{\rm min}\, (n,\tilde n)}\left(
\sum_{v_\ell=0}^{{\rm min}\,(\sigma_{(\ell)},\tau_{(\ell)})}(-1)^{v_\ell}
\ell^{v_\ell}v_\ell!\pmatrix{\sigma_{(\ell)}\cr v_\ell\cr}
\pmatrix{\tau_{(\ell)}\cr v_\ell\cr}\frac{1}{N^{2v_\ell}}\right).
\eqn\eeswaba$$
In this expression we recognize our expression for $\sigma_{n,\tilde n}
(P)$ that summarizes the contributions of different components of the 
moduli spaces of minimal-area maps to the partition function of the 
topological rigid string.  Because of the remarkable generating formula 
for the Euler numbers of the moduli spaces, \eecm , we can also identify 
the combinatorial factors in front of the sum over all homotopies $a_j,b_j$ 
in \eeqe\ as the Euler numbers of the moduli spaces.  Hence, the partition 
function of the large-$N$ QCD in two dimensions at $\lambda=0$, as given 
by \eeqe , is equal to the partition function of the topological rigid 
string as summarized by eqn.~\eess , assuming we set $g_{\rm string}=1/N$.  
This is one of the central results of this paper.  

\section{Non-Zero Target Area/Non-Zero String Tension}

At non-zero area/non-zero couping constant, the full results of [\grta] 
can also be interpreted in simple geometrical terms.  For $\su N$ Yang-Mills 
theory, the large-$N$ expansion gives 
$$\eqalign{\CZ(G,&\lambda A,N)=\sum_{n,\tilde n}\frac{1}{n!\tilde n!}
\e{-(n+\tilde n)\lambda A/2}\sum_{s,t}(-1)^s\frac{(\lambda A)^{s+t}}{s!t!}
N^{(n+\tilde n)(2-2G)-s-2t}\frac{(n-\tilde n)^{2t}}{2^t}\cr
&{}\times\sum_{k=0}^{\infty}(-1)^k\left(\matrix{2G+k-3\cr k\cr}\right)
\sum_{p_1,\ldots p_s\in T_2}\sum_{a_1,b_1,\ldots a_G,b_G}\delta(p_1\ldots
p_s\tilde\Omega_{n,\tilde n}^k\prod_{j=1}^Ga_jb_ja_j^{-1}b_j^{-1}).\cr}
\eqn\eedbaaa$$
Following [\grta], we decompose $(n-\tilde n)^2$ into $-2n\tilde n+[n(n-1)+
\tilde n(\tilde n-1)] +[n+\tilde n]$, and interpret the terms as coming from 
simple orientation-reversing connecting necks, simple orientation-preserving 
connecting tubes, and collapsed handles respectively.  $s$ in \eedbaaa\ 
counts the number of simple branchpoints that contribute to the area 
dependence of the partition function.  

We have already interpreted the $\lambda=0$ part of the partition function 
as a calculation of the Euler number of certain moduli spaces of minimal-area 
maps, and can be consequently written as an integral over the moduli space 
of the exponential of a specific two-form, cf.~\eezxzx .  It can be 
straightforwardly shown that the whole partition function at non-zero 
$\lambda$ can also be written as an integral of a specific form over the 
same moduli space.  In order to get a better insight into the situation, 
consider first the following integral 
$$\int_{\CM_{G,k}}\e{-\CR-\lambda\CW}\eqn\eedl$$
on the moduli spaces $\CM_{G,k}\equiv\tilde\CM_{G,k}/S_k$, with 
$\tilde\CM_{G,k}\equiv(M_G)^k$.  Here we have defined two-forms $\CR$ and 
$\CW$ on the moduli spaces by 
$$\CR=\sum_{i=1}^k\CR_i,\qquad\CW=\sum_{i=1}^k\CW_i,\eqn\eedm$$
where $\CR_i$ and $\CW_i$ are the curvature two-form and the volume two-form 
on the $i$-th copy of $M_G$.  On $\tilde\CM_{G,k}$, the integral gives
$$\int_{\tilde\CM_{G,k}}\e{-\CR-\lambda\CW}=\frac{(-1)^k}{k!}\int_{\tilde
\CM_{G,k}}(\CR+\lambda\CW)^k=\frac{(-1)^k}{k!}\sum_{s=0}^k\pmatrix{k\cr s\cr}
\lambda^s\int_{\tilde\CM_{G,k}}\CW^s\wedge\CR^{k-s}.\eqn\eedn$$
Since $\tilde\CM_{G,k}$ is a direct product of $k$ copies of $M_G$, this 
expression can be further reduced to 
$$\eqalign{\int_{\tilde\CM_{G,k}}&\e{-\CR-\lambda\CW}=\frac{(-1)^k}{k!}
\sum_{s=0}^k\pmatrix{k\cr s\cr}\lambda^s\int_{\tilde\CM_{G,k}}\sum_{i_1\ldots
i_s}\CW_{i_1}\wedge\ldots\CW_{i_s}\wedge\sum_{j_1\ldots j_{k-s}}\CR_{j_1}
\wedge\ldots\CR_{j_{k-s}}\cr
&=\frac{(-1)^k}{k!}\sum_{s=0}^k\pmatrix{k\cr s\cr}\lambda^s\,k!
\int_{\tilde\CM_{G,k}}\CW_1\wedge\ldots\wedge\CW_s\wedge\CR_{s+1}\wedge\ldots
\wedge\CR_k\cr
&=(-1)^k\sum_{s=0}^k\pmatrix{k\cr s\cr}\lambda^s\int_{\tilde\CM_{G,s}}\CW_1
\wedge\ldots\wedge\CW_s\int_{\tilde\CM_{G,k-s}}\CR_1\wedge\ldots\wedge
\CR_{k-s}\cr
&\qquad\qquad=(-1)^k\sum_{s=0}^k\pmatrix{k\cr s\cr}\lambda^s\;{\rm Vol}\,
(\tilde\CM_{G,s})\;\chi(\tilde\CM_{G,k-s}).\cr}\eqn\eedo$$
Our moduli spaces $\CM_{G,k}$ are factors of $\tilde\CM_{G,k}$ by $S_k$, and 
an analogous evaluation of the integral leads to
$$\int_{\CM_{G,k}}\e{-\CR-\lambda\CW}=(-1)^k\sum_{s=0}^k\lambda^s\;{\rm Vol}\,
(\CM_{G,s})\;\chi(\CM_{G,k-s}).\eqn\eedp$$
This is exactly the area dependence encountered in \eedbaaa , since ${\rm 
Vol}\,(\CM_{G,s})=A^s/s!$.  

To facilitate our further discussion, it is useful to place the theory with 
non-zero $\lambda$ into a wider context.  It is indeed well known that in two 
dimensions, one can deform the Yang-Mills Lagrangian by an infinite number 
of new terms, and write
$$\CL'=\int_M\phi F+\int_M\der^2x\rtg f(\phi),\eqn\eefff$$
where $f(\phi)$ is an arbitrary class function on the Lie algebra of the 
Yang-Mills gauge group $\CG$, and can be written as a sum over the infinite 
number of Casimir operators of $\CG$.   In the context of the topological 
interpretation of the Yang-Mills theory as summarized in \S{1.2}, these new 
terms correspond to the higher BRST cohomology classes expressed in terms 
of the ghost-for-ghost field $\phi$.  We can still interpret the partition 
function of the deformed Yang-Mills theory \eefff\ as a correlation function 
of the BRST cohomology classes in the underlying topological Yang-Mills 
theory (cf.\ \eeac ):  
$$\left\langle 1\right\rangle_{f(\phi)}=\left\langle\exp\left\{
-\int_M\left(\phi F-\psi\wedge\psi\right)-\int_M\rtg f(\phi)\right\}
\right\rangle_{\rm topo.\ YM}\eqn\eededede$$
At large $N$, all the new terms in the Yang-Mills Lagrangian have a 
corresponding string interpretation, discussed in [\genqcd].  

In the topological rigid string theory, these deformed partition functions 
can be reconstructed as follows.  The moduli spaces of minimal-area maps, as 
discussed in the analysis of the topological rigid string theory in \S{5}, 
carry a natural cohomological structure.  In particular, they carry an 
infinite number of natural cohomology classes, which are the analogy of the 
``stable'' or ``universal'' cohomology classes that can be naturally defined 
on moduli spaces of other moduli problems studied in the literature 
[\topograv].  Recall that we have parametrized our moduli spaces by the 
target locations $P_1,\ldots,P_s$ of the moduli of minimal-area maps.  The 
fixed metric on the target manifold $M$ induces a natural induced-area 
two-form $\CW(P_i)$ on $\CM$ for each $P_i$, the only non-zero components of 
$\CW(P_i)$ being along the direction of $P$.  The natural cohomology classes 
are then generated by specific two-forms on $\CM$, which are in one-to-one 
correspondence with the elements of the set of all conjugacy classes of all 
possible moduli (``modulus types'' from now on).  Given a fixed modulus type 
$\alpha$, for example a branchpoint of degree $p$, define a two-form 
$\CO_\alpha$ on $\CM$ as the sum of the induced-area two-forms $\CW(P_i)$ 
where $i$ runs over all moduli in the conjugacy class of $\alpha$.  

In addition to these natural cohomology classes $\CO_\alpha$ on $\CM$ which 
are all two-forms, another cohomology class is needed to establish a relation 
with the partition functions of the generalized Yang-Mills theory at large 
$N$.  This class is a zero-form on $\CM$, whose value in each point of the 
moduli space is equal to the induced area of the worldsheet of the 
corresponding minimal-area map.  We will denote this cohomology class by 
$\CO_0$.  

By analogy with the Yang-Mills formula \eededede , we now claim that the 
large-$N$ expansion of the partiton functions in the generalized Yang-Mills 
theory can be written as integrals of the exponential of a linear combination 
of the natural cohomology classes, combined with the density on the moduli 
spaces that is already present in the theory at $f(\phi)\equiv 0$, 
$$\CZ(G,f(\phi),N)=\int_\CM\e{-\CR_0(\CM)-c_0\CO_0-\sum c_\alpha\CO_\alpha}
\eqn\eererere$$
More precisely, a natural map $\Upsilon$ exists that associates with each 
choice of $f(\phi)$ in the generalized Yang-Mills theory a linear combination 
of the natural cohomology classes $\CO_0,\CO_\alpha$ of the moduli spaces, 
such that formula \eererere\ is valid.  The results of [\genqcd] can be 
considered a direct verification of this statement.  For a given $f(\phi)$, 
the specific coefficients $c_0,c_\alpha$ of $\Upsilon(f(\phi))\equiv 
c_0\CO_0+c_\alpha\CO_\alpha$ can be directly inferred from [\genqcd].  

Hence, in the string representation, any coupling constant of the generalized 
Yang-Mills theory multiplies a linear combination of the natural cohomology 
classes of the moduli spaces, and the large-$N$ expansion of the partition 
function of the generalized Yang-Mills theory can be written as a correlation 
function of the corresponding cohomology classes in the topological rigid 
string.  In particular, the $\lambda$ term of the standard Yang-Mills 
Lagrangian is interpreted in the topological rigid string theory as a 
specific linear combination of the natural cohomology classes of the moduli 
spaces,
$$\Upsilon(\lambda\phi^2)=\lambda\left(\CO_0+2\CO_b+2\CO_r-2\CO_p-
\CO_h\right).\eqn\eeswqc$$
Here $\CO_0$ is the zero-form that has been defined above, while the 
remaining contributions come from the natural two-forms $\CO_\alpha$: $\CO_b$ 
is the natural two-form that corresponds to the conjugacy class of a simple 
branchpoint, $\CO_r$ corresponds to the simple orientation-reversing 
collapsed neck, $\CO_p$ to the orientation-preserving collapsed tube, and 
$\CO_h$ to the collapsed handle.  The last three contributions would be 
missing if we change the gauge group from $\su N$ to $\u N$.  

Several remarks are in order:

\item{1.} Even though in the $\u N$ theory the moduli that correspond to 
collapsed handles do not contribute to either the partition function at zero 
$\lambda$ or to the theory where $\lambda$ is the only non-zero coupling in 
$f(\phi )$, collapsed handles do emerge when we consider higher Casimirs in 
$f(\phi)$ (see [\genqcd]).  Hence, collapsed handles are not specifics of the 
$\su N$ theory, and we cannot get rid of them by restriction to $\u N$.  It 
is an advantage of our formulation of the QCD string theory over possible 
alternative formulations in terms of holomorphic maps that collapsed handles 
emerge as simple moduli from the outset.  

\item{2.}  In the topological rigid string theory, the difference between 
$\su N$ and $\u N$  with only $\lambda$ non-zero seems to be a matter of 
choice of the specific combination of the cohomology classes in \eererere , 
and neither choice seems to be particularly singled out.  

An explicit field representation of the universal cohomology classes 
in the topological rigid string would require a detailed analysis of the 
equivariant BRST cohomology of the topological rigid string and is not an 
easy task, but we can at least discuss the simplest observables that emerge 
in the theory at non-zero $\lambda$.  

Notice first that if we integrate out the auxiliary field $B^\mu$ in the 
topological rigid string theory defined by the Lagrangian \eebub , the 
on-shell BRST algebra is given (in the specific gauge \eebua ) by 

$$\eqalign{[Q,x^\mu]&=\psi^\mu,\qquad\{Q,\psi^\mu\}=0,\cr
\vphantom{\int}\{Q,\chi^\mu\}&=-\frac{1}{2a}(1+a\psi\chi)\Delta 
x^\mu+\frac{1}{2}\chi^\mu(\nabla\psi\cdot\p x)-
\frac{1}{2}\psi^\mu(\nabla\chi\cdot\p x)\cr
&\qquad\qquad\qquad\qquad{}+\left(\delta^\mu_\nu-\p_ax^\mu h^{ab}\p_b 
x^\lambda g_{\lambda\nu}\right)\christ\nu\sigma\rho\chi^\sigma\psi^\rho.\cr}
\eqn\eedg$$
Define now
$$\delta\CL=\lambda\ints\rth\left\{1+2a\psi\ast\chi\right\}.\eqn\eepepe$$
In two dimensions, one can demonstrate by a direct calculation that 
$\delta\CL$ is on-shell BRST invariant up to gauge transformation of the 
worldsheet gauge symmetries, i.e.
$$[Q,\delta\CL]={\rm pure\ gauge}.\eqn\eedh$$
This is all we need to be able to evaluate the deformed partition function, 
$$\CZ(\lambda)=\int\e{-\CL-\delta\CL},\eqn\eedf$$
perturbatively in $\lambda$.  

The first term in $\delta\CL$ is just the bosonic Nambu-Goto Lagrangian 
(i.e.\ the induced area term), and $\lambda$ is the string tension.  The 
fermionic term in $\delta\CL$ is an improvement that makes $\delta\CL$ an 
admissible deformation from the point of view of the topological BRST 
symmetry.  

We claim that $\delta\CL$ is a field-theoretical representation of the 
cohomology class $\CO_0$ defined above.  To show this, we evaluate the 
deformed partition function \eedf\ perturbatively in $\lambda$.  Because of 
the nice BRST properties of the Lagrangian, we can evaluate the deformed 
path integral semiclassically.  The part with no fermions is the bosonic 
Nambu-Goto action and measures the total induced area of the worldsheet, 
which on the moduli spaces of minimal area maps reduces to
$$\lambda\ints\rth=\lambda(n+\tilde n)A,\eqn\eedk$$
with $n$ and $\tilde n$ being the number of sheets in the orientation 
preserving and orientation reversing sectors respectively.  $\lambda$ is 
indeed the string tension.  The term linear in quantum fluctuations 
$\delta x^\mu$ is proportional to the first variation of the Nambu-Goto 
action, and vanishes on the moduli spaces of minimal-area maps.  The terms 
quadratic in quantum fluctuations could only affect the one-loop determinants 
in target dimensions higher than two.  Hence, the one-loop determinants still 
cancel each other (up to a possible sign), and $\delta\CL$ indeed reduces 
to the evaluation of the induced area of the worldsheet in the minimal-area 
map, which is the definition of the cohomology class $\CO_0$.  

In the topological rigid string theory defined by the deformed Lagrangian 
\eebv , it is natural to consider 
$$\delta\CL'=\lambda\ints\rth\left\{1+\psi\ast\chi+(\psi\ast\chi)^2\right\}.
\eqn\eedc$$
Notice that $\delta\CL'$ is related to the Lagrangian $\CL$ of the 
topological rigid string as defined in \eebv\ by a double-commutator formula,
$$\CL=\{Q,[\bar Q,\ints\frac{\rth}{1-\psi\ast\chi}]\}=\{Q,[\bar Q,
\frac{1}{\lambda}\delta\CL']\}.\eqn\eede$$
This relation is not accidental, but its more thorough explanation would lead 
us beyond topological theory [\future].  Thus, we will only show that the 
deforming term $\delta\CL'$ is weakly BRST invariant, which also allows us to 
use it as a deformation of the Lagrangian.  On the moduli space of 
minimal-area maps, where $\psi^\mu=\chi^\mu=0$ and $\Delta x^\mu=0$, the 
Lagrangian $\CL$ vanishes, and the double commutator formula \eede\ reduces to
$$\{\bar Q,[Q,\delta\CL']\}=0.\eqn\eedi$$
Hence, on the moduli space, the new term in the Lagrangian is BRST invariant 
up to a $\bar Q$-closed term,
$$[Q,\delta\CL']=\Gamma,\qquad \{\bar Q,\Gamma\}=0,\eqn\eedj$$
which again allows us to add $\delta\CL'$ to the Lagrangian of the 
topological rigid string and evaluate the deformed path integral 
semiclassically.  The only difference from the previous calculation comes 
from the zero-mode integration, where we are left with the four-fermi term 
$(\psi\ast\chi)^2$ that accompanies the curvature-dependent four-fermi term 
of the topological rigid string Lagrangian in the saturation of the zero mode 
integral in the fermi sector.  Effectively, this changes the measure on the 
moduli spaces, from the Euler measure to a measure that depends on the volume 
element of the components of the moduli spaces, according to formula 
\eererere .  Consequently, $\delta\CL'$ is a field-theoretical representation 
of a certain linear combination of the cohomology classes $\CO_0$ and 
$\CO_\alpha$. 

The whole dependence of the partition functions on all coupling constants of 
the generalized Yang-Mills theory would require a detailed information about 
the field-theoretical realization of the natural cohomology classes of the 
moduli spaces of minimal-area maps introduced above, as well as a more 
detailed technical understanding of some puzzling aspects of the path 
integral of the theory (in particular, in the degenerate homotopy classes).  
In general, we should not expect the full structure of observables and 
correlation functions of the topological rigid string theory to be much 
simpler that that of conventional topological string theory [\topograv].  
Already at this stage, however, we can claim that the simplest deforming term 
corresponds to the Nambu-Goto induced-area term, improved by fermionic terms 
in order to become a BRST invariant observable.  For generic coupling 
constants, the topological rigid string interpretation of the large-$N$ 
partition functions follows from the cohomological formula \eererere .  

\endpage
\chapter{Conclusions}

In this paper, we have presented a topological rigid string theory, and 
discussed it in two dimensions as a theory of QCD strings.  Since an 
extension of the theory to higher dimensions will be discussed elsewhere 
[\future], we limit our conclusions to several remarks on those aspects of 
the two-dimensional theory that were not discussed in the body of this 
paper.  

(1)  Although we focused out attention on partition functions of the 
topological rigid string and their relation to the large-$N$ expansion of 
the QCD partition functions, our results can be easily extended to 
correlation functions of Wilson loops in arbitrary representations of 
$\su N$.  It is a straightforward exercise to show that the Wilson loop 
correlation functions at $\lambda=0$ calculate the Euler numbers of moduli 
spaces of minimal-area maps from worldsheets with boundaries to the targets 
with Wilson loops, while the area dependence emerges from the volume of the 
moduli spaces, exactly as in the case of the partition functions.  The only 
difference from the calculation of the partition functions is in the slightly 
more complicated geometry of the moduli spaces of such minimal-area maps.  

(2)  Instead of $\su N$ Yang-Mills 
theory, one can study the large-$N$ expansion for alternative series of gauge 
groups, $\so N$ and $\sp N$.  The results of [\grta] have been extended to 
these alternative cases in [\othergrs], leading to theories of unoriented 
closed strings (as expected).  We will argue now that the corresponding 
string theories can also be described by a topological rigid string theory, 
as follows.  We have seen in \S{2.5} that for harmonic topological sigma 
models, a canonical orbifold theory exists; when extended to the topological 
rigid string, this orbifold construction makes closed strings unoriented, and 
introduces open strings (with the standard, Neumann boundary conditions on 
both ends) as twisted states.  We claim that this orbifold theory describes 
the large-$N$ string theory of QCD with the alternative gauge groups.  (The 
difference between $\so N$ and $\sp N$ corresponds to the sign choice in the 
definition of the $\ztwo$ orbifold action on the closed string sector).  This 
conjecture raises an obvious question, as we certainly do not expect open 
strings in the large-$N$ expansion of QCD Yang-Mills theory without matter, 
for either gauge group.  This apparent paradox has a surprising resolution:  
The orbifold version of the topological rigid string theory does indeed 
contain an open string sector, with Neumann boundary conditions of both ends 
of open strings; all open-string states are however unphysical, since they 
are all homotopically trivial.  

(3) As a next logical step, one can try to couple the string to dynamical 
quarks (perhaps by a choice of worldsheet boundary conditions that break the 
worldsheet $\CN=2$ supersymmetry of the topological rigid string down to 
$\CN=1$ along the boundary), and compare the results to the known properties 
of the 't~Hooft model [\thmodel].  The existence of a natural coupling 
between the strings and quarks would serve as an independent check on the 
validity of our string theory as a two dimensional QCD string theory.  

(4)  Worldsheet and spacetime duality.  2D QCD string theory contains 
essentially two coupling constants: the string tension $\lambda$ (or more 
precisely, its dimensionless version $\hat\lambda\equiv\lambda A$), and the 
string coupling constant $g_{\rm string}\equiv 1/N$.  Results of [\dualqcd] 
suggest that the theory exhibits interesting duality properties under 
$\hat\lambda\rightarrow\hat\lambda^{-1}$ (at least on the torus).  In the 
spacetime Yang-Mills theory, such a duality would represent a strong-weak 
coupling duality (S-duality), while in the string theory it would correspond 
to a T-duality (since $\lambda$ is a worldsheet coupling constant).  We can 
also speculate about dualities that would mix $\hat\lambda$ and the string 
coupling constant $1/N$.  Indeed, dualities of this type (i.e.\ dualities 
that interchange the rank of the gauge group with the gauge coupling 
constant) are not unknown in low-dimensional gauge theories [\rankdual].  In 
the string representation, these dualities would mix a worldsheet coupling 
constant with the string coupling constant, and would be an example of what 
has come to be called U-duality [\stringdual].  In QCD string theory, a 
U-duality interchanging $\hat\lambda$ and $N$ would map the large-$N$ string 
theory to a Wilson-like  strong-coupling string theory [\wilsonstr], thus 
leading to two alternative string descriptions of a given Yang-Mills gauge 
theory.  

(5)  With a Lagrangian formulation of the QCD string theory, one can write 
down its corresponding string field theory, following the standard lore 
[\barton].  Such a QCD string field theory should be equivalent to the 
Yang-Mills theory, although not manifestly so.  Since the spacetime 
Yang-Mills theory is completely solved, 2D QCD might represent an excellent 
opportunity to test concepts of string field theory.  Thus, one can study the 
relatively well-understood string non-perturbative effects in 2D QCD (such as 
the Douglas-Kazakov phase transition and the finite-$N$ effects [\finiten]) 
from the point of view of string theory, and even look for their possible 
worldsheet interpretation (cf.\ the recent ideas of Green and Polchinski, 
[\polchgr]).  It would also be very instructive to see how spacetime gauge 
invariance and the Yang-Mills field emerge in string field theory.  Any 
progress in that direction would be helpful in the search for a microscopic 
derivation of the QCD string theory.  
\bigskip
\centerline{\fourteenrm Acknowledgements}
\medskip
It is a pleasure to thank M.~Douglas, D.~Gross, Y.~Nambu, A.~Polyakov, 
I.~Singer, L.~Susskind, W.~Taylor and E.~Witten for discussions at various 
stages of this work.  I am grateful to the Aspen Center for Physics, where 
a part of this work was done, for its hospitality.  This work has been 
supported in part by NSF Grant No.\ PHY90-21984.  
\refout
\end